\PassOptionsToPackage{longend}{algorithm2e}
\documentclass{theoretics}

\usepackage{bm} 
\usepackage{cancel}
\usepackage{amsfonts}
\usepackage{textcomp}
\renewcommand{\copyright}{\textcopyright}

\usepackage[colorinlistoftodos]{todonotes}

\usepackage{xspace}

\usepackage{tikz}
\usetikzlibrary{calc}
\usetikzlibrary{shapes,decorations.markings,arrows.meta,fit}

\usepackage[normalem]{ulem}

\usepackage{relsize}

\newcommand{\norm}[1]{\|#1\|}
\newcommand{\set}[1]{\{#1\}}                    
\newcommand{\setof}[2]{\{{#1}\mid{#2}\}}        
\newcommand{\dom}{\textsf{Dom}}

\newcommand{\degree}{\text{\sf deg}}

\newcommand{\opt}{\text{\sf opt}}

\newcommand{\inner}[1]{\langle #1 \rangle}

\newcommand{\calM}{\mathcal M}
\newcommand{\calS}{\mathcal S}

\newcommand{\calD}{\mathcal D}
\newcommand{\calL}{\mathcal L}

\newcommand{\calT}{\mathcal T}
\newcommand{\calZ}{\mathcal Z}

\newcommand{\defeq}{\stackrel{\text{def}}{=}}

\newcommand{\R}{\mathbb R} 
\newcommand{\Rp}{{\mathbb R}_{\tiny +}} 

\newcommand{\cd}{\text{ :- }}

\newcommand{\bigjoin}{\mathlarger{\mathlarger{\mathlarger{\Join}}}}

\newcommand{\inn}{\text{\sf in}}
\newcommand{\out}{\text{\sf out}}
\newcommand{\vars}{\text{\sf vars}}
\newcommand{\nodes}{\text{\sf nodes}}

\newcommand{\mon}{\text{\sf MON}}
\newcommand{\sub}{\text{\sf SUB}}

\newcommand{\agm}{\textsf{AGM}\xspace}
\newcommand{\panda}{\textsf{PANDA}\xspace}
\newcommand{\csma}{\textsf{CSMA}\xspace}

\newcommand{\fhtw}{\textsf{fhtw}}
\newcommand{\subm}{\textsf{subw}}

\usepackage{afterpage}

\definecolor{@ThCSdarkred}{RGB}{195,30,30}
\definecolor{@ThCSdarkblue}{RGB}{50,70,120}

\title{\panda: Query Evaluation in Submodular Width}

\ThCSauthor[relai]{Mahmoud Abo Khamis}{mahmoudabo@gmail.com}[0000-0003-3894-6494]
\ThCSauthor[relai]{Hung Q. Ngo}{hung.q.ngo@gmail.com}[0000-0002-3498-4744]
\ThCSauthor[washington]{Dan Suciu}{suciu@cs.washington.edu}[0000-0002-4144-0868]

\ThCSaffil[relai]{Relational AI, Berkeley, USA}
\ThCSaffil[washington]{University of Washington, Seattle, USA}

\ThCSthanks{A preliminary version of this article appeared at PODS 2017~\cite{DBLP:conf/pods/Khamis0S17}.

  Part of this work was conducted while the authors participated in the
  Fall 2023 {\em Simons Program on Logic and Algorithms in Databases and AI.}
  We thank the anonymous reviewers for many insightful comments.}

\ThCSshortnames{M.\ Abo Khamis, H.Q.\ Ngo, D.\ Suciu}  
\ThCSshorttitle{\panda: Query Evaluation in Submodular Width}
\ThCSyear{2025}
\ThCSarticlenum{12}
\ThCSreceived{Jun 5, 2024}
\ThCSrevised{Jan 7, 2025}
\ThCSaccepted{Feb 28, 2025}
\ThCSpublished{Apr 30, 2025}
\ThCSdoicreatedtrue
\ThCSkeywords{Shannon inequalities, submodular width, disjunctive datalog, query evaluation, degree constraints}

\begin{document}

\maketitle

\begin{abstract}
    In recent years, several information-theoretic upper bounds have been introduced on the
    output size and evaluation cost of database join queries. These bounds vary in their
    power depending on both the type of statistics on input relations and the query plans
    that they support.  This motivated the search for algorithms that can compute the output
    of a join query in times that are bounded by the corresponding information-theoretic
    bounds.  In this paper, we describe $\panda$, an algorithm that takes a
    Shannon-inequality that underlies the bound, and translates each proof step into an
    algorithmic step corresponding to some database operation.  \panda computes answers to a
    conjunctive query in time given by the submodular width plus the output size of the
    query. The version in this paper represents a significant simplification of the original
    version in~\cite{DBLP:conf/pods/Khamis0S17}.
\end{abstract}

\section{Introduction}
Answering {\em conjunctive queries} efficiently is a fundamental problem in the theory and
practice of database management, graph algorithms, logic, constraint satisfaction, and
graphical model inference, among
others~\cite{DBLP:conf/pods/KhamisNR16,DBLP:conf/pods/GottlobGLS16,DBLP:journals/talg/GroheM14,DBLP:books/aw/AbiteboulHV95}.
In a {\em full conjunctive query}, the input is a set of relations (or tables, or
constraints), each with a set of attributes (or variables), and the task is to list all {\em
satisfying assignments}, i.e.,~assignments to all variables that simultaneously satisfy all
input relations.  Each such assignment is called an {\em output tuple}, and their number is
the {\em output size}. For example, the query
\begin{align}
Q(a,b,c) \cd E(a,b) \wedge E(b,c) \wedge E(c,a) \label{eqn:triangle}
\end{align}
is a full conjunctive query asking for the list of all triangles in a (directed) graph with
edge relation $E$.
In contrast, in a {\em Boolean conjunctive query}, we only ask whether one such
assignment exists. The query
\[ Q() \cd E(a,b) \wedge E(b,c) \wedge E(c,a) \]
asks whether there is a triangle in the graph.
More generally, in a proper {\em conjunctive query}, any subset of variables can occur in
the head of the query. These are called {\em free variables}. For example, the query
\[ Q(a) \cd E(a,b) \wedge E(b,c) \wedge E(c,a) \]
asks for the list of all nodes $a$ that are part of a triangle.
Other variants of the query evaluation problem include counting output
tuples, performing some aggregation over them, or enumerating them under some restrictions.

In the case of a full conjunctive query, the runtime of any algorithm is at least as large
as the size of the output.  This has motivated the study of upper bounds on the sizes of the
query outputs. The corresponding graph-theoretic problem is to bound the number of
homomorphic images of a small graph within a larger graph; this problem has a long history
in graph theory and isoperimetric
inequalities~\cite{MR0031538,MR599482,MR859293,MR1338683,10.1145/3196959.3196990,MR1639767}.
One such bound is the {\em \agm bound}, which is a formula that, given only the
cardinalities of the input relations, returns an upper bound on the output size of the
query~\cite{DBLP:journals/siamcomp/AtseriasGM13}. Moreover, the bound is tight, meaning that
there exist relations of the given cardinalities where the query's output has size equal to
the \agm bound (up to a query-dependent factor).  This immediately implies that no algorithm
can run in time lower than the \agm bound in the worst-case. Thus, an algorithm that runs in
this time is called a {\em Worst-Case Optimal Join}
algorithm.
The \agm bound for
query~\eqref{eqn:triangle} is $O(|E|^{3/2})$; algorithms for listing all triangles within
this amount of time has been known for decades~\cite{DBLP:conf/stoc/Itai77}.
For general full conjunctive queries, Grohe and Marx~\cite{DBLP:conf/soda/GroheM06}, and
Atserias, Grohe, and Marx~\cite{DBLP:journals/siamcomp/AtseriasGM13,DBLP:conf/focs/AtseriasGM08}
devised a join-project query plan that can compute the output of a full conjunctive query
to within a linear factor of the \agm bound, which is very close to optimality.
They also showed that a join-only query plan cannot achieve this bound.
A few years later, a new class of join algorithms, not based on join and project operators,
was able to achieve the \agm bound~\cite{DBLP:conf/icdt/Veldhuizen14,DBLP:journals/jacm/NgoPRR18}
and thus achieve worst-case optimality.

In practice, especially in database systems, the input cardinalities are not sufficient to
model what we know about the data, and definitely not sufficient to predict how good or bad
a query plan is. Other data characteristics such as functional dependencies and distinct
value counts are often collected and used to optimize queries~\cite{DBLP:conf/icdt/000122,DBLP:journals/pvldb/LeisGMBK015};
furthermore, practical queries often have ``relations'' that are infinite in pure
cardinalities. For example, the output size of this query
\[ Q(a,b) \cd R(a) \wedge S(b) \wedge a+b=10 \] is obviously at most $\min\{|R|,|S|\}$, but
the \agm bound is $|R|\cdot |S|$. The relation $a+b=10$ has infinite cardinality. There
has thus been a line of research to strengthen the \agm bound to account for increasingly
finer classes of input statistics.
Specifically, Gottlob, Lee, Valiant and Valiant~\cite{DBLP:journals/jacm/GottlobLVV12,DBLP:conf/pods/GottlobLV09}
applied the entropy argument to derive a bound on the output size of a full conjunctive query
under general functional dependencies. Their bound, generalizing the \agm-bound, is an optimization
problem whose constraints are the Shannon inequalities.
This idea was a seed for a series of works that extended the entropy argument to account for
finer classes of constraints, including degree constraints~\cite{DBLP:conf/pods/KhamisNS16,
DBLP:conf/pods/Khamis0S17,10.1145/3651597}, and
$\ell_p$-norm bounds on degree sequences~\cite{10.1145/3651597}.

Designing WCOJ algorithms that match these stronger
bounds becomes harder since the bounds under more constraints are tighter. An algorithm
called  \csma is described in~\cite{DBLP:conf/pods/KhamisNS16}, which only accounts for
relation cardinalities and functional dependencies.

In database management systems jargon, a WCOJ algorithm is a {\em multi-way join operator};
this operator, for some classes of queries, is asymptotically faster than the traditional
binary join operator~\cite{DBLP:journals/jacm/NgoPRR18}. Given a complicated conjunctive query,
however, before applying a WCOJ algorithm, it is often beneficial to come up with a {\em
query plan} that decomposes the query into simpler subqueries. Query plans guided by
(hyper)tree decompositions have proved to be very useful both in theory and in
practice~\cite{DBLP:conf/pods/GottlobGLS16}. In particular, query plans represented by a
single tree decomposition can be used to answer a Boolean conjunctive query or a full
conjunctive query in time bounded by $O(N^{\fhtw}+|\text{output}|)$, where $\fhtw$ is
the {\em fractional hypertree width}~\cite{DBLP:journals/talg/GroheM14} of the query, and
$|\text{output}|$ is the size of the output. This runtime is sensitive to the output size,
and thus it is more adaptive than a single WCOJ application.\footnote{It should be noted,
however, that the sub-queries of the plan are still being answered with a WCOJ operator.}
For example, using tree-decomposition-based query plans, we can identify whether a large
graph with $N$ edges contains the homomorphic images of a $k$-cycle in time $O(N^2)$
(assuming a constant $k$).

Motivated by finding the fixed-parameter tractability boundary for Boolean conjunctive
queries, or equivalently {\em constraint satisfaction problems}, under the regime of
unbounded-arity inputs, Marx~\cite{DBLP:journals/jacm/Marx13} came up with a beautifully
novel idea: instead of fixing a single query plan (i.e.,~tree-decomposition)
up front, we can consider multiple query
plans, and partition the data to make use of different plans for different parts of the
data. The improved complexity measure that Marx introduced is called the {\em submodular
width}, $\subm$, which is always less than or equal to $\fhtw$.

The $\subm$ algorithm from Marx~\cite{DBLP:journals/jacm/Marx13} has some limitations. First,
it assumes all input relations have the same cardinality $N$; in particular, it is not known
how to define the width and design similar algorithms under functional dependencies or degree
constraints. Second, the runtime of the
algorithm is {\em not} $O(N^{\subm})$, but a polynomial in this quantity.
Third, the $\subm$-notion and the algorithm were not defined for general conjunctive queries
with arbitrary free variables.

\textbf{Contributions.}
This paper connects all three of these lines of research: WCOJ algorithms, fractional
hypertree width, and submodular width into a single framework, while dealing with arbitrary
input degree constraints (which is a superset of functional dependencies and cardinality
constraints), {\em and} arbitrary conjunctive queries. The bridge that connects these
algorithms and concepts is information theory. In particular, our contributions include the
following:

We show how a generalized version of the classic {\em Shearer's inequality}~\cite{MR1639767}
can be used to derive upper bounds on the output size of a {\em disjunctive datalog rule}
(DDR), which is a generalization of a conjunctive query.
The upper bound is a generalization of the \agm bound to include degree constraints and
DDRs. The introduction of DDR and its information theoretic output size bound in studying
conjunctive queries is our first major technical novelty.
DDRs are interesting in their own right. They form the building blocks of
\emph{disjunctive datalog}~\cite{DBLP:journals/tods/EiterGM97}, which is a significant
extension of datalog.  Disjunctive datalog has a long history: it emerged in logic
programming~\cite{DBLP:books/daglib/0069187,DBLP:journals/amai/Minker94}, and is used for
knowledge representation, representing incomplete information, and constraint satisfaction.
Our bound on the output size of a DDR represents a bound on the output size of the
\emph{minimal model of the disjunctive datalog program} consisting of that single rule.

Next, we show that certain symbolic manipulations of the information inequality can be converted
into a query evaluation plan for the DDR that runs in time bounded by the predicted upper
bound.
This idea of converting a proof of an information inequality into a query evaluation plan
is our second major technical novelty.
The algorithm is called \panda,
which stands for ``{\bf P}roof-{\bf A}ssisted e{\bf N}tropic {\bf D}egree-{\bf A}ware''.
In particular, \panda is worst-case optimal for
DDRs under arbitrary degree constraints. Even when restricted to only conjunctive queries,
this is already beyond the state of the art in WCOJ algorithms because previous WCOJ
algorithms cannot meet the tightened bounds under degree constraints.

Lastly, we explain how to define the notions of $\fhtw$ and $\subm$ under degree constraints
and for conjunctive queries with arbitrary free variables.
We show how \panda can be used to answer arbitrary conjunctive queries with arbitrary
degree constraints, in time bounded by $\tilde O(N^{\subm} + |\text{output}|)$, where $\tilde O$ hides a polylogarithmic factor in $N$.
These results close the gaps left by Marx' work.
For example, with \panda,
the $k$-cycle query can now be answered in $\tilde O(N^{2-1/\lceil k/2\rceil })$ time, which is
sub-quadratic, and matches the specialized cycle detection algorithm
from Alon, Yuster, and Zwick~\cite{DBLP:journals/algorithmica/AlonYZ97}.

The results in this paper were first announced in a conference paper~\cite{DBLP:conf/pods/Khamis0S17}.  The current paper makes several
significant improvements:
\begin{itemize}
\item In~\cite{DBLP:conf/pods/Khamis0S17}, we used both the primal and the dual linear
  program to guide \panda: the primal gives an optimal polymatroid $\bm h$, while the dual
  represents the basic Shannon inequalities.  In the new version, we use only the dual,
  which significantly simplifies the algorithm.  The algorithm is described only in terms of
  an information inequality and its proof (called a {\em witness}), which correspond
  precisely to a feasible solution to the dual program.  We only need to describe the primal
  and dual programs later, in Sec.~\ref{sec:subw}, where we introduce the degree-aware
  submodular width, which is defined in terms of the primal.

\item Previously, we needed a {\em proof sequence} to drive the algorithm; it was difficult
  to prove that a proof sequence exists; for example, no proof sequence existed in our
  earlier framework~\cite{DBLP:conf/pods/KhamisNS16}.  In the new version, we describe
  \panda\ without the need for an explicit proof sequence, which again simplifies it. If
  needed, a proof sequence can still be extracted from the new version of the algorithm.

\item One difficulty in the earlier presentation of \panda was the
  need to recompute the proof sequence after a reset step.  This is no
  longer necessary here.
\end{itemize}

{\bf Paper Outline} This paper is organized as follows. Section~\ref{sec:prelim} presents
background concepts needed to understand the techniques and results of the paper; in
particular, it introduces {\em disjunctive datalog rules} (DDR), a generalization of
conjunctive queries, reviews necessary background on {\em information theory},  and defines
the class of statistics on input relations that the \panda algorithm supports, which are
called {\em degree constraints}. Section~\ref{sec:info:inequality} discusses the class of
information inequalities that are at the center of our work, where they are used to
both derive upper bounds on the output size of a query and guide the \panda algorithm.
Section~\ref{sec:main:results} states the main algorithmic result, which says that the
$\panda$ algorithm meets this information-theoretic upper bound {\em if} the bound is a
Shannon inequality, i.e.,~it does not involve non-Shannon
inequalities~\cite{DBLP:journals/tit/ZhangY97,zhang1998characterization}.
Section~\ref{sec:algorithm} presents the core $\panda$ algorithm, which constructs a
step-by-step proof of the Shannon inequality, and converts each step into a database
operation. Section~\ref{sec:subw} defines the degree-constraint aware submodular width, and
shows how to use disjunctive datalog rules to compute a conjunctive query in time given by
the submodular width.  We conclude in Section~\ref{sec:conclusion}.

\section{Preliminaries}
\label{sec:prelim}

\subsection{Database instances and conjunctive queries (CQ)}
\label{subsec:conjunctive}

Fix a set $\bm V$ of variables (or attributes).  An {\em atom} is an expression of the form
$R(\bm X)$ where $R$ is a relation name and $\bm X \subseteq \bm V$ is a set of attributes.
A {\em schema}, $\Sigma$, is a set of atoms.  We shall assume throughout that distinct atoms
in $\Sigma$ have distinct attribute sets. If $R$ is a relation name in $\Sigma$, we write
$\vars(R)$ for its attribute set, and define $\vars(\Sigma)\defeq \bm V$ to be the
set of all attributes in the schema $\Sigma$.

Given a countably infinite domain $\dom$, we use $\dom^{\bm X}$ to denote the set of tuples
with attributes $\bm X \subseteq \bm V$. A {\em $\Sigma$-instance} is a map $D$ that assigns
to each relation name $R$ in $\Sigma$ a finite subset $R^D \subseteq\dom^{\vars(R)}$.
Technically, we should use $\Sigma^D \defeq (R^D)_{R \in \Sigma}$ to denote the
$\Sigma$-instance; however, to reduce notational burden, instead of writing $R^D$ and
$\Sigma^D$, we will often write $R$ and $\Sigma$ when the instance is clear from the context.
Given $\bm X \subseteq \bm V$ and a tuple $\bm t \in \dom^{\bm V}$, we write $\pi_{\bm
X}(\bm t)$ to denote the {\em projection} of $\bm t$ onto the variables $\bm X$.

The {\em full natural join} (or {\em full join} for short) of the $\Sigma$-instance is the set of tuples
$\bm t \in \dom^{\bm V}$ that satisfy all atoms in $\Sigma$:
\begin{align}
  \bigjoin \Sigma & \defeq \setof{\bm t \in \dom^{\bm V}}{
    \pi_{\vars(R)}(\bm t) \in R,
    \forall R \in \Sigma
  }
  \label{eq:full:join}
\end{align}
This set of tuples is sometimes called in the literature the {\em universal table} of the
instance $\Sigma$.

Given a schema $\Sigma$,  a {\em conjunctive query} is the expression
\begin{align}
    Q(\bm F) &\cd  \bigwedge_{R(\bm X)\in \Sigma} R(\bm X) \label{eq:cq}
\end{align}
where $\bm F \subseteq \bm V$ is called the set of {\em free variables},
and $Q(\bm F)$ is the {\em head atom} of the query.
Atoms in $\Sigma$ are called the {\em body atoms} of the query.
The output $Q(\bm F)$ of an input instance $\Sigma$ is the projection of the
full join~\eqref{eq:full:join} onto the free variables $\bm F$:
$Q(\bm F) \defeq \pi_{\bm F}(\bigjoin \Sigma)$.

When $\bm F = \bm V$, we call the query a {\em full conjunctive query}.
When $\bm F = \emptyset$, we call the query a {\em Boolean conjunctive query}, whose answer
$Q()$ is either {\sf true} or {\sf false}, and it is true if and only if the full
join~\eqref{eq:full:join} is non-empty.

Our complexity results are expressed in terms of
{\em data complexity}; in particular, we consider the number of atoms and variables to be a constant, i.e.,~$|\Sigma| + |\bm V| = O(1)$,
and the complexity is a function of the instance size. We
define the {\em size} of a $\Sigma$-instance as:
\begin{align}
    \norm{\Sigma}\defeq \sum_{R(\bm X) \in \Sigma} |R|.
    \label{eqn:input:size}
\end{align}
The notation $\norm{\Sigma}$ is used in part to distinguish it from $|\Sigma|$ which
counts the number of atoms in~$\Sigma$.

\subsection{Tree decompositions and free-connex queries}
\label{subsec:td}

Consider a conjunctive query in the form~\eqref{eq:cq}. A {\em tree decomposition} of $Q$ is
a pair $(T,\chi)$, where $T$ is a tree and $\chi: \nodes(T) \rightarrow 2^{\vars(Q)}$ is a
map from the nodes of $T$ to subsets of $\vars(Q)$ that satisfies the following properties: for all atoms $R(\bm X)$ in $\Sigma$,
there is a node $t \in \nodes(T)$ such that $\bm X \subseteq \chi(t)$; and, for any variable
$X \in \vars(Q)$, the set $\setof{t}{X \in \chi(t)}$ forms a connected subtree of $T$. Each
set $\chi(t)$ is called a {\em bag} of the tree-decomposition, and we will assume w.l.o.g.
that the bags are distinct, i.e.,~$\chi(t)\neq\chi(t')$ when $t\neq t'$ are nodes in $T$.

A {\em free-connex} tree decomposition for $Q$ is a tree-decomposition for $Q$ with an
additional property that there is a connected subtree $T'$ of $T$ for which
$\bm F = \bigcup_{t \in \nodes(T')} \chi(t)$.
The query $Q$ is {\em free-connex acyclic} iff there is a free-connex tree decomposition
$(T, \chi)$ in which every bag is covered by an input atom;
namely, for every $t \in \nodes(T)$, there exists an input atom $R(\bm X) \in \Sigma$
where $\chi(t) \subseteq \bm X$.

The following result is
well-known~\cite{DBLP:conf/vldb/Yannakakis81,DBLP:conf/csl/BaganDG07}.

\begin{lemma}
    If $Q$ is a free-connex acyclic conjunctive query of the form~\eqref{eq:cq}, then we can compute
    its output in time $\tilde O(\norm{\Sigma} + |Q(\bm F)|)$.
    In particular, after a preprocessing time of $\tilde O(\norm{\Sigma})$, we can list the output tuples one by one with constant-delay between them.
    \label{lmm:free:connex:acyclic}
\end{lemma}

In particular, for this class of queries, the runtime is the best one can hope for: input
size plus output size.

\subsection{Disjunctive Datalog rules (DDR)}
\label{subsec:disjunctive}

{\em Disjunctive Datalog rules}~\cite{DBLP:journals/tods/EiterGM97} are a
generalization of conjunctive queries, where the head of the query
can be a disjunction, formalized as follows.
Let $\Sigma_{\inn}$ and $\Sigma_{\out}$ be two schemas,
called input and output schema respectively.  We associate to these two schemas the
following \emph{disjunctive Datalog rule} (DDR)
\begin{align}
  \bigvee_{Q(\bm Z)\in \Sigma_{\out}} Q(\bm Z) &\cd \bigwedge_{R(\bm X)\in\Sigma_{\inn}} R(\bm X) \label{eq:disjunctive:rule}
\end{align}
The DDR is uniquely defined by the two schemas
$\Sigma_{\inn}$ and $\Sigma_{\out}$.  The syntax
in~\eqref{eq:disjunctive:rule} does not add any new information, but
is intended to be suggestive for the following semantics:

\begin{definition} \label{def:feasible} Let $\Sigma_{\inn}$ be an input instance.  A {\em
model} (or {\em feasible output}) for the rule~\eqref{eq:disjunctive:rule} is an instance
$\Sigma_{\out}$, such that the following condition holds:
for every tuple $\bm t \in \bigjoin \Sigma_{\inn}$, there is an output atom
$Q(\bm Z) \in \Sigma_{\out}$ for which $\pi_{\bm Z}(\bm t) \in Q$.
Similar to~\eqref{eqn:input:size},
the {\em size} of the output instance $\Sigma_{\out}$ is defined as
\begin{align}
\norm{\Sigma_{\out}} \defeq \sum_{Q(\bm Z) \in \Sigma_{\out}} |Q|. \label{eqn:ddr:size}
\end{align}
A model is {\em minimal} if we cannot remove a tuple from any output relation without
violating the feasibility condition.
\end{definition}

In English, a feasible output needs to store each tuple
$\bm t \in \bigjoin \Sigma_{\inn}$ in at least one of the output
relations $Q(\bm Z) \in \Sigma_{\out}$.  The query evaluation problem
is to compute a minimal model. Note that a conjunctive query is a disjunctive Datalog rule
where $\Sigma_{\out}$ has a single atom.

DDRs are interesting. We illustrate the concept here with a few examples.

\begin{example}
  Consider the following DDR,
  \begin{align*}
    Q(X,Z) &\cd R(X,Y) \wedge S(Y,Z)
  \end{align*}
  where $\Sigma_\inn=\set{R(X,Y), S(Y,Z)}$ and
  $\Sigma_\out=\set{Q(X,Z)}$.  A model (or feasible output) to the DDR
is any superset of $\pi_{XZ}(R \Join S)$.
  Consider now the following DDR:
  \begin{align*}
    A(X) \vee B(Y) &\cd R(X,Y)
  \end{align*}
  One model is $A := \pi_X(R)$, $B := \emptyset$.  Another
  model is $A:= \emptyset$, $B := \pi_Y(R)$. Both are minimal.
  Many other models exist.

  A non-trivial DDR is the following:
  \begin{align}
    A(X,Y,Z) \vee B(Y,Z,W) &\cd R(X,Y) \wedge S(Y,Z) \wedge U(Z,W) \label{eq:ab:query}
  \end{align}
  To compute the rule, for each tuple $(x,y,z,w)$ in the full join, we
  must either insert $(x,y,z)$ into $A$, or insert $(y,z,w)$ into $B$.
  The output size is the larger of the resulting relations $A$ and $B$.
  We shall see later in the paper that, for this rule, there is a model of size
  $O(\sqrt{|R| \cdot |S| \cdot |U|})$, which is a non-trivial result.
\end{example}

\subsection{Entropic vectors and polymatroids}
\label{subsec:entropy}

For general background on information theory and polymatroids, we refer the reader
to~\cite{Yeung:2008:ITN:1457455}.
Given a discrete (tuple of) random variable(s) $\bm X$ over a domain $\dom(\bm X)$ with
distribution $p$, the (Shannon)
{\em entropy} of $\bm X$ is defined as
\[
  h(\bm X) \defeq -\sum_{\bm x\in \dom(\bm X)} p(\bm X=\bm x) \log p(\bm X=\bm x)
\]
The {\em support} of the distribution is the set of all $\bm x$ where $p(\bm X=\bm x) > 0$.
We will only work with distributions of finite support. Let $N$ be the support size, then it
is known that $0 \leq h(\bm X) \leq \log N$. The upper bound follows from the concavity of
$h$ and Jensen's inequality. Moreover, $h(\bm X) = 0$ iff $X$ is {\em deterministic}, i.e.,~it has a singleton support, and $h(\bm X) = \log N$ iff $X$ is {\em uniform}, meaning that
$p(\bm X=\bm x) = 1/N$ for all $\bm x$ in the support.

If $\bm V$ is a set of jointly distributed random variables, then the vector
$\bm h = (h(\bm X))_{\bm X \subseteq \bm V}\in\Rp^{2^{\bm V}}$ is called an
{\em entropic vector}.\footnote{Given two sets $A$ and $B$, we use
$B^A$ to denote the set of all functions $f:A\rightarrow B$.}
Implicitly, $h(\emptyset) = 0$; thus, this vector is actually of dimension $2^{|\bm V|}-1$.
We will often write $\bm X\bm Y$ for the union $\bm X \cup \bm Y$.
In particular, $h(\bm X \bm Y) = h(\bm X \cup\bm Y)$.

Starting with Shannon himself, the study of linear functions on entropic vectors has been a
central topic in information theory. The two basic linear functions are defined by the
so-called information measures. An {\em information measure} is an expression $\mu = (\bm
Y|\bm X)$ or $\sigma = (\bm Y; \bm Z|\bm X)$, where $\bm X, \bm Y, \bm Z$ are disjoint
subsets of the set of variables $\bm V$.  We call $\mu$ a {\em monotonicity} and $\sigma$ a
{\em submodularity} information measure respectively. A monotonicity measure $\mu = (\bm
Y|\bm X)$ is called {\em unconditional} iff $\bm X = \emptyset$. Similarly, a submodularity
measure $\sigma = (\bm Y;\bm Z|\bm X)$ is called {\em unconditional} iff $\bm X =
\emptyset$.
For any vector $\bm h \in
\Rp^{2^{\bm V}}$, we define the linear functions:
\begin{align*}
  h(\mu) & \defeq h(\bm X\bm Y) - h(\bm X) && \mu=(\bm Y|\bm X)\\
  h(\sigma) & \defeq h(\bm X\bm Y)+h(\bm X\bm Z)-h(\bm X)-h(\bm X\bm Y\bm Z)
    && \sigma=(\bm Y;\bm Z|\bm X)
\end{align*}

We write $\mon$ for the set of monotonicity measures, i.e.,~the set of all $\mu = (\bm Y|\bm X)$
where $\bm X,\bm Y\subseteq \bm V$ are disjoint.
Similarly, we write $\sub$ for the
set of submodularity measures.

A \emph{polymatroid} is a vector $\bm h$
that satisfies the {\em basic Shannon inequalities}:
\begin{align}
  h(\emptyset) &= 0,
  & \forall \mu \in \mon,\ \ h(\mu) &\geq 0,
  &\forall \sigma \in \sub,\ \ h(\sigma) & \geq 0
  \label{eq:shannon}
\end{align}
The latter two are called monotonicity and submodularity constraints respectively.
Every entropic vector is a polymatroid, but the converse is not
true~\cite{DBLP:journals/tit/ZhangY97}.
A {\em Shannon inequality} is a linear inequality that is derived from the
basic Shannon inequalities; equivalently, a linear inequality is a
Shannon inequality iff it is satisfied by all polymatroids.

{\bf Discussion} In the literature, the basic Shannon inequalities are
often restricted to {\em elemental} ones, which are submodularity
measures of the form $(A;B|\bm X)$ and monotonicity measures of the
form $(A|\bm V-\set{A})$, where $A,B \in \bm V$ are single variables,
and $\bm X\subseteq \bm V-\set{A,B}$. The reason is that the elemental
inequalities are sufficient to prove every Shannon inequality; for
example $h(B;CD|A) = h(B;C|A)+h(B;D|AC) \geq 0$, when both
$h(B;C|A)\geq 0$ and $h(B;D|AC)\geq 0$.
Given $m \defeq |\bm V|$, the total number of elemental
basic Shannon inequalities is $m(m-1)2^{m-3}+m$.
\footnote{Specifically, there are
${m \choose 2}$ ways to choose $A$ and $B$ and $2^{m-2}$ ways to choose $\bm X$
resulting in
${m \choose 2}\cdot 2^{m-2}$ elemental submodularities.
Additionally, there are $m$ elemental monotonicities.}
However, for the
purpose of the \panda algorithm, it is preferable to consider all basic
Shannon inequalities, because this may lead to a smaller exponent of
the polylog factor of the algorithm, as we will see in Section~\ref{sec:algorithm}.

\subsection{Statistics on the data}
\label{subsec:degrees}

In typical database engines, various statistics about the data are maintained and used for
cardinality estimation, query optimization, and other purposes. Common statistics include
the number of distinct values in a column, the number of tuples in a relation, and
functional dependencies. A robust abstraction called ``degree constraints'' was introduced
in~\cite{DBLP:conf/pods/Khamis0S17} that captures many of these statistics. This section
provides a formal definition of degree constraints, which are also the constraints that
$\panda$ can support.

Let $\delta = (\bm Y|\bm X)$ be a monotonicity measure,
$\Sigma$ be a database instance with schema $\Sigma$,
$R \in \Sigma$ be a relation in this instance,
and $\bm x \in \dom^{\bm X}$ be a tuple with schema $\bm X$.
We define the quantity {\em degree} of $\bm x$ with respect to $\bm Y$ in $R$, denoted by
$\degree_{R}(\bm Y | \bm X = \bm x)$, as follows:
\begin{itemize}
  \item When both $\bm X \subseteq \vars(R)$ and $\bm Y \subseteq \vars(R)$,
then $\degree_{R}(\bm Y|\bm X = \bm x)$ is the number of times the given
$\bm X$-tuple $\bm x$ occurs in $\pi_{\bm X\bm Y}(R)$.
  \item When $\bm X$ and $\bm Y$ are arbitrary with respect to $\vars(R)$, then we define
the restriction of $R$ to $\bm X \cup \bm Y$ to be the relation
$R'(\bm X\bm Y) := \dom^{\bm X \cup \bm Y} \ltimes R$,\footnote{$S \ltimes T$ denotes
the {\em semi-join reduce} operator defined by $S \ltimes T \defeq \pi_{\vars(S)}(S \Join T)$.} and set
  \[
\degree_{R}(\bm Y | \bm X = \bm x) \defeq \degree_{R'}(\bm Y | \bm X = \bm x)
  \]
\end{itemize}
Finally, define the {\em degree} of the monotonicity measure $\delta = (\bm Y|\bm X)$ in $R$,
denoted by $\degree_{R}(\delta)$, to~be
\begin{align}
    \degree_{R}(\bm Y|\bm X) \defeq & \max_{\bm x \in \dom^{\bm X}} \degree_{R}(\bm Y|\bm X = \bm x) \label{eq:degree:delta}
\end{align}
Note that, for infinite $\dom$, if $\bm Y \not\subseteq \vars(R)$, then $\degree_{R}(\bm Y|\bm X)=\infty$, unless
$R=\emptyset$, in which case $\degree_{R}(\bm Y|\bm X)=0$. We say that $R$ is a {\em guard}
of $\delta = (\bm Y|\bm X)$ if $\bm Y \subseteq \vars(R)$, and note that, if $R \neq
\emptyset$, then $R$ is a guard of $\delta$ iff $\degree_{R}(\bm Y|\bm X)< \infty$.

If $\bm X=\emptyset$ and $\bm Y = \vars(R)$, then the degree is the {\em cardinality} of
$R$, $\degree_{R}(\bm Y|\emptyset) = |R|$. If $\deg_R(\delta)=1$ then there is a {\em
functional dependency} in $R$ from $\bm X$ to $\bm Y$. If the number of unique values in a
column $A$ of $R$ is $k$, then $\deg_R(A|\emptyset)=k$. Given a schema instance $\Sigma$,
define the degree of $\delta = (\bm Y|\bm X)$ in the instance $\Sigma$ as:
\begin{align}
  \degree_{\Sigma}(\delta) &\defeq \min_{R \in \Sigma}\degree_{R}(\delta).
  \label{eqn:degree:sigma}
\end{align}

Let $\Delta\subseteq\mon$ be a set of monotonicity measures and
$\bm N:\Delta \rightarrow \Rp$ be numerical values for each
$\delta \in\Delta$. Throughout this paper, we write $N_\delta$ instead of $N(\delta)$, and
define $n_\delta \defeq \log N_\delta$ for all $\delta \in\Delta$.  We
view the pair $(\Delta, \bm N)$ as a set of {\em degree constraints},
and we say that an instance $\Sigma$ {\em
  satisfies} the degree constraints $(\Delta, \bm N)$ iff
$\degree_{\Sigma}(\delta) \leq N_\delta$ for all
$\delta \in \Delta$.  In that case, we write~$\Sigma \models (\Delta, \bm N)$.

\section{On a Class Information Inequalities}
\label{sec:info:inequality}

The entropy argument and Shearer's lemma in particular~\cite{MR859293} is a powerful
information-theoretic tool in extremal combinatorics~\cite{entropy-counting}. Friedgut and
Kahn~\cite{MR1639767} applied the argument to bound the number of homomorphic
copies of a graph
in a larger graph; this is a special case of the full conjunctive query problem.
Grohe and Marx~\cite{DBLP:conf/soda/GroheM06}, with further elaboration
in~\cite{DBLP:conf/focs/AtseriasGM08}
showed how Shearer's lemma can be used to bound the output size of a full CQ given input
cardinality constraints.
Briefly, let $\Sigma$ be the input schema of a full CQ of the form~\eqref{eq:cq},
\begin{align}
    Q(\bm V) &\cd  \bigwedge_{R(\bm X)\in \Sigma}R(\bm X). \label{eq:full:cq}
\end{align}
where the head atom $Q(\bm V)$ has all the variables.
Given any non-negative weight vector
$\bm w = (w_{\bm X})_{R(\bm X)\in \Sigma}$
that forms a fractional edge cover of the hypergraph $(\bm V, \bm E)$ where $\bm E := \{\bm X \mid R(\bm
X)\in \Sigma\}$,
the output size of the full CQ is bounded by
\begin{align}
    |Q| &\leq \prod_{R(\bm X)\in \Sigma} |R|^{w_{\bm X}}.
    \label{eqn:agm}
\end{align}
This is known the AGM-bound~\cite{DBLP:journals/siamcomp/AtseriasGM13}.
The bound is a direct consequence of {\em Shearer's inequality}~\cite{MR859293}, which states
that the inequality
\begin{align}
    h(\bm V) &\leq  \sum_{R(\bm X)\in \Sigma} w_{\bm X} \cdot h(\bm X), \label{eqn:shearer}
\end{align}
holds for every entropic vector $\bm h \in \R_+^{2^{\bm V}}$ if and only if the weights $\bm w$
form a fractional edge cover of the hypergraph $(\bm V, \bm E)$ defined above.

The above results can only deal with cardinality constraints of input relations, and if the
input query is a conjunctive query. We extend these results to disjunctive Datalog rules and
handle general degree constraints. We start in the next section with a generalization of
Shearer's inequality and show how that implies an output cardinality bound for DDRs. In
later sections, we use the information inequality to drive the \panda algorithm.

\subsection{Size Bound for DDRs from Information Inequalities}
\label{subsec:upperbound}

This section develops an information-theoretic bound for the output size of a DDR under
general degree constraints. Inequality~\eqref{eqn:ddr:shearer} below is a generalization of
Shearer's inequality~\eqref{eqn:shearer}, and the bound~\eqref{eq:q:inequality} is a
generalization of the AGM bound~\eqref{eqn:agm} for DDRs and general degree constraints. (Recall
the notation for the size of a model $\Sigma_{\out}$ in~\eqref{eqn:ddr:size}.) In what
follows, for a given schema $\Sigma$ and an atom $R(\bm X) \in \Sigma$, for brevity we will
also write $\bm X \in \Sigma$ as we assume a one-to-one correspondence between atoms and
their variables.

\begin{theorem} \label{th:upper:bound}
Consider a DDR of the form~\eqref{eq:disjunctive:rule}
with input and output schemas $\Sigma_{\inn}$ and $\Sigma_{\out}$ respectively. Let
$\Delta\subseteq \mon$ be a set of monotonicity measures.
Suppose that there exist two {\em non-negative} weight vectors $\bm w := (w_\delta)_{\delta \in \Delta}$ and
$\bm \lambda := (\lambda_{\bm Z})_{\bm Z \in\Sigma_{\out}}$
with $\norm{\bm \lambda}_1=1$,
where the following inequality holds for all entropic vectors $\bm h$:
\begin{align}
    \sum_{\bm Z \in \Sigma_{\out}}\lambda_{\bm Z}\cdot h(\bm Z)
    & \leq \sum_{\delta \in \Delta} w_\delta \cdot h(\delta)\label{eqn:ddr:shearer}
\end{align}
  Then, for any input instance $\Sigma_{\inn}$ for the DDR~\eqref{eq:disjunctive:rule},
  there exists a model $\Sigma_{\out}$ for the DDR that satisfies: (Recall that $|\Sigma_\out|$ is the number of atoms in $\Sigma_\out$.)
  \begin{align}
    \norm{\Sigma_{\out}} &\leq
        |\Sigma_\out|\cdot
      \prod_{\delta \in \Delta} \left(\degree_{\Sigma_{\inn}}(\delta)\right)^{w_\delta}
      \label{eq:q:inequality}
  \end{align}
\end{theorem}
\begin{proof} The plan is to use the entropy argument~\cite{MR859293},
  where we define a uniform probability distribution on a certain subset $\bar Q \subseteq
  \bigjoin \Sigma_{\inn}$, denote $\bm h$ its entropic vector, then
  use~\eqref{eqn:ddr:shearer} to prove~\eqref{eq:q:inequality}.

  Let $\bm V \defeq \vars(\Sigma_{\inn})$ be the set of all input variables. Notice that,
  for any joint distribution on $\bm V$ with entropic vector $\bm h$, we have $h(\delta)
  \leq \log \degree_{\Sigma_{\inn}}(\delta)$ for all $\delta \in \Delta$. This is trivially
  true if $\delta \in \Delta$ is not guarded
  by any relation in $\Sigma_{\inn}$
  (see Sec~\ref{subsec:degrees} for the notion of guardedness).
  Otherwise, the inequality follows from the fact that the uniform
  distribution on a finite support has the maximum entropy:
  \begin{align*}
    h(\delta) = h(\bm Y | \bm X) &=
    \sum_{x \in \dom^{\bm X}} p(\bm X = \bm x) h(\bm Y | \bm X = \bm x)\\
    &\leq
    \sum_{x \in \dom^{\bm X}} p(\bm X = \bm x) \log \degree_{\Sigma_{\inn}}(\bm Y | \bm X = \bm x)\\
    &\leq
    \sum_{x \in \dom^{\bm X}} p(\bm X = \bm x) \log \degree_{\Sigma_{\inn}}(\delta)\\
    &= \log \degree_{\Sigma_{\inn}}(\delta)
  \end{align*}

  Next, we construct both the set $\bar Q$ and a model $\Sigma_{\out}$ for the DDR as follows.
  Initially, set $\bar Q = \emptyset$, and $Q = \emptyset$ for all $Q \in \Sigma_{\out}$.
  Iterate over the tuples $\bm t \in \bigjoin \Sigma_{\inn}$ in some arbitrary order. For
  each $\bm t$: if $\exists Q \in \Sigma_{\out}$ s.t. $\pi_{\vars(Q)}(\bm t) \in Q$, then
  ignore $\bm t$; otherwise,
  insert $\bm t$ into $\bar Q$ and
  insert $\pi_{\vars(Q)}(\bm t)$ into $Q$ for every $Q \in \Sigma_{\out}$.
  In the end, we have constructed a model
  $\Sigma_{\out}$ for the DDR.
  Furthermore, $|\bar Q| = \norm{\Sigma_{\out}}$.

  Finally, consider the uniform distribution on $\bar Q$, i.e.,~the distribution where each
  tuple in $\bar Q$ is chosen randomly with probability $1/|\bar Q|$. Let $\bm h$ be the entropy
  vector of this distribution. Notice that for each $Q \in \Sigma_{\out}$, the marginal
  distribution on $\vars(Q)$ is also uniform, because $|Q| = |\bar Q|$; in particular, $h(\bm
  Z) = \log |Q| = \log |\bar Q|$ for all $Q(\bm Z) \in \Sigma_{\out}$. Then,
  noting that $\norm{\bm \lambda}_1=1$, the following holds:
  \begin{align*}
    \log \frac{\norm{\Sigma_{\out}}}{|\Sigma_\out|} = \log |\bar Q|
    = \sum_{\bm Z\in \Sigma_{\out}} \lambda_{\bm Z} \cdot h(\bm Z) \leq \sum_{\delta \in \Delta} w_\delta\cdot  h(\delta)
        \leq \sum_{\delta \in \Delta} w_\delta \cdot \log \degree_{\Sigma_{\inn}}(\delta) 
  \end{align*}
\end{proof}

In order to obtain the best bound, we need to choose the weights $\bm w$ and $\bm \lambda$
to minimize quantity $\prod_{\delta \in \Delta} \left(\degree_{\Sigma_{\inn}}(\delta)\right)^{w_\delta}$ on the right-hand side of~\eqref{eq:q:inequality}. Specifically, we want to
minimize the linear objective
\begin{align}
  \min_{\bm \lambda, \bm w} \sum_{\delta \in \Delta} w_\delta \cdot \log \degree_{\Sigma_{\inn}}(\delta)
  \label{eqn:log:objective}
\end{align}
subject to the constraints that $\bm w\geq 0$, $\bm \lambda \geq 0$, $\norm{\bm \lambda}_1=1$,
and that inequality~\eqref{eqn:ddr:shearer} holds for all entropic vectors $\bm h$.

For general monotonicity measure $\Delta$, it is an open problem to characterize
the weight vectors $\bm w$ and $\bm \lambda$ for which~\eqref{eqn:ddr:shearer} holds
for all entropic vectors $\bm h$.
In particular, the difficulty is related to the problem of
characterizing the entropic region in information theory~\cite{Yeung:2008:ITN:1457455}.
Hence, to make the problem tractable, we relax the upper bound by requiring~\eqref{eqn:ddr:shearer}
to hold for all polymatroids $\bm h$.

\begin{definition}[Polymatroid bound]
The following bound is called the {\em polymatroid bound} for disjunctive datalog rules:
\begin{align}
  b^*_{\Delta,\bm N} &\defeq \min \sum_{\delta \in \Delta} w_\delta \cdot \log \degree_{\Sigma_{\inn}}(\delta) \label{eqn:polymatroid:bound}\\
  \text{subject to:} &\qquad \norm{\bm \lambda}_1=1 \nonumber \\
  &\qquad \text{inequality } \eqref{eqn:ddr:shearer} \text{ holds for all polymatroids } \bm h \in \Rp^{2^{\bm V}} \label{eqn:linear:constraint}\\
  &\qquad \bm w \geq 0, \bm \lambda \geq 0 \nonumber
\end{align}
\label{defn:polymatroid:bound}
\end{definition}
Note that the bound is stated in log-scale, so that the objective is linear. The constraint~\eqref{eqn:linear:constraint}
is a linear constraint, as explained below. In particular, the polymatroid bound is a linear program.

\begin{proposition}
The constraint that~\eqref{eqn:ddr:shearer} holds for all polymatroids $\bm h$ is a linear constraint in the
weight vectors $\bm w$ and $\bm \lambda$ and auxiliary variables.
\end{proposition}
\begin{proof}
  In the space $\R^{2^{\bm V}}$, the set of polymatroids is a polyhedron of the form
  $\{ \bm h \mid \bm A \bm h \geq \bm 0 \}$ for an appropriately defined matrix $\bm A$.
  Inequality~\eqref{eqn:ddr:shearer} is a linear inequality of the form
  $\bm b^\top \bm h \geq \bm 0$, where $\bm b$ is a linear function of the weights $\bm w$ and $\bm \lambda$.
  From the Gale-Kuhn-Tucker variant of Farkas' lemma\footnote{\url{https://en.wikipedia.org/wiki/Farkas\%27_lemma}},
  inequality~\eqref{eqn:ddr:shearer} holds for all polymatroids $\bm h$ if and only if
  there is no $\bm h$ for which $\bm b^\top \bm h < 0$ and $\bm A\bm h \geq \bm 0$,
  which holds if and only if there is a vector $\bm x$ such that
  $\bm A^\top \bm x = \bm b$ and $\bm x \geq \bm 0$.
  The last condition is a linear constraint in the weights $\bm w$ and $\bm \lambda$, and
  in the dual variables $\bm x$.
\end{proof}

    The following proposition says that the polymatroid bound linear program always has a rational solution.
    Therefore, given any degree constraints $(\Delta, \bm N)$ and vectors $(\bm\lambda,\bm w)$ that define a valid Shannon inequality~\eqref{eqn:ddr:shearer},
    there are always rational vectors $(\bm\lambda^*,\bm w^*)$ that also define a valid Shannon inequality and satisfy:
    \begin{align*}
        \sum_{\delta\in\Delta}w^*_\delta\cdot \log \degree_{\Sigma_\inn}(\delta) \leq
        \sum_{\delta\in\Delta}w_\delta\cdot \log \degree_{\Sigma_\inn}(\delta)
    \end{align*}
    \begin{proposition}
        Given any degree constraints $(\Delta, \bm N)$, the polymatroid bound linear program from Eq.~\eqref{eqn:polymatroid:bound} has an optimal solution $(\bm\lambda^*,\bm w^*)$ which is rational and independent of $(\Delta, \bm N)$.
        \label{prop:rational:shannon}
    \end{proposition}
    \begin{proof}
        The constraints of the linear program have integer coefficients.
        Moreover, these coefficients are independent of the given statistics $(\Delta, \bm N)$.
        (The statistics are only used in the objective function.)
        Consider the polytope defined by the constraints of the linear program.
        No matter what the objective function is, there is always an optimal solution which is a vertex of this polytope. By Cramer's rule, all these vertices are rational.
    \end{proof}

The question of how tight the polymatroid bound is (and its entropic counterpart) has
an intriguing connection to information theory. We refer the reader to~\cite{DBLP:conf/pods/Khamis0S17,10175769}
for more in-depth discussions and results.

\subsection{Equivalent Formulations of Inequality~\eqref{eqn:ddr:shearer}}
\label{subsec:equivalent}

Shearer's result~\cite{MR859293} states that inequality~\eqref{eqn:shearer} holds for all
polymatroids $\bm h$ iff the weights form a fractional edge cover of a certain hypergraph.
This section shows an analogous characterization for the
generalization~\eqref{eqn:ddr:shearer}, and states an ``integral'' version of this
characterization that shall be used by the \panda algorithm.
The following lemma is a variant of Farkas' lemma~\cite{schrijver-book} applied to our specific setting.

\begin{lemma}
Let $\bm a \in \R^{2^{\bm V}}$ be a coefficient vector. The following inequality is a
Shannon inequality:
\begin{align}
  \sum_{\bm X \subseteq \bm V} a_{\bm X}h(\bm X) \geq & 0 \label{eq:general:inequality}
\end{align}
if and of only if there exist non-negative coefficients
$\bm m = (m_\mu)_{\mu \in \mon}$ and $\bm s = (s_\sigma)_{\sigma \in \sub}$
such that the following equality holds as an identity over $2^{|\bm V|}$ symbolic variables
$h(\bm X)$, $\bm X \subseteq \bm V$:
\begin{align}
\sum_{\bm X \subseteq \bm V} a_{\bm X}h(\bm X|\emptyset)
  &= \sum_{\mu \in \mon} m_\mu h(\mu) + \sum_{\sigma \in \sub} s_\sigma  h(\sigma)
  \label{eq:witness}
\end{align}
We call the tuple $(\bm m, \bm s)$
a {\em witness} of~\eqref{eq:general:inequality}.
If $\bm a$ is rational, then there exists a rational witness.
Furthermore, $\bm m$ and $\bm s$ are a function of $\bm a$ and $|\bm V|$.
\label{lmm:witness}
\end{lemma}
\begin{proof}
Clearly if~\eqref{eq:witness} holds as an identity, then~\eqref{eq:general:inequality}
follows trivially.
The converse is a direct consequence of a variant of Farkas' lemma, in particular
Corollary 7.1h in Schrijver's classic text~\cite{schrijver-book}, which states:
if the polyhedron $P = \{ \bm x \; | \; \bm A\bm x \geq \bm b\}$ is not empty,
and if inequality $\inner{\bm c, \bm x} \geq d$ holds for all $\bm x \in P$,
then there exists $d'\geq d$ for which $\inner{\bm c, \bm x} \geq d'$
is a non-negative linear combination of the inequalities in $\bm A\bm x
\geq \bm b$.\footnote{We thank the anonymous reviewer for pointing out this
specific variant that helps simplify our proof.}

In our context, $P$ is the polyhedron defined by~\eqref{eq:shannon}
and $\bm b = \bm 0$, and the inequality is defined by  $\bm c = \bm a$ and
$d=0$.
Farkas' lemma thus implies that if~\eqref{eq:general:inequality}
is a Shannon inequality, then $\inner{\bm a, \bm h} \geq 0$ is a non-negative linear combination
of the Shannon inequalities in~\eqref{eq:shannon}, namely there
are coefficients $\bm m,\bm s,e^+,e^-$ such that
the following identity holds
over the variables $h(\bm X)$,~$\bm X \subseteq \bm V$:
\begin{align}
\inner{\bm a, \bm h} = \sum_{\mu \in \mon} m_\mu h(\mu) + \sum_{\sigma \in \sub} s_\sigma h(\sigma)
+(e^+ - e^-)h(\emptyset).
\label{eqn:a:h}
\end{align}
where $e^+$ and $e^-$ are coefficients associated with $h(\emptyset) \geq 0$
and $-h(\emptyset) \geq 0$.
Setting $h(\bm X)=1$ for all $\bm X$ in~\eqref{eqn:a:h}, we conclude that $\inner{\bm a,\bm 1}=e^+ - e^-$,
and thus~\eqref{eq:witness} holds.

The fact that $\bm m$ and $\bm s$ are a function of $\bm a$ and $|\bm V|$ (and the bounds
on their representation sizes) can be found in Chapter 10 of Schrijver's book~\cite{schrijver-book}.
\end{proof}

In words, Lemma~\ref{lmm:witness} says that the LHS of~\eqref{eq:general:inequality} is a
positive linear combination of monotonicity and submodularity measures. From the lemma, it is
not hard to show that Shearer's inequality~\eqref{eqn:shearer} holds whenever the weights
form a fractional edge cover of the corresponding hypergraph.
Given a Shannon inequality of the form~\eqref{eq:general:inequality},
the witness $(\bm m, \bm s)$
from Lemma~\ref{lmm:witness} can be computed by solving the following linear program:
The variables are $m_\mu$ and $s_\sigma$, which are non-negative.
The constraints ensure that Eq.~\eqref{eq:witness} is an identity.
Specifically, for each $\bm X \subseteq \bm V$, there is a constraint
that says that $a_{\bm X}$ must be equal to the weighted sum of terms $m_\mu$ and $s_\sigma$ that contribute to the coefficient of $h(\bm X)$ on the RHS of Eq.~\eqref{eq:witness}.
The linear program has no objective. Instead, we just compute a feasible solution.

To guide the $\panda$ algorithm later, we will need an ``integral'' version of the lemma above.
Given a set $S$, a multiset $\calS$ {\em over} $S$ is a multiset whose members are in $S$. The
size of a multiset $\calS$, denoted by $|\calS|$, is the number of its members, counting
multiplicity.
If the coefficients $\bm \lambda$ and $\bm w$ in Eq.~\eqref{eqn:ddr:shearer} are rational, then the above linear program has a rational solution
$(\bm m, \bm s)$, hence
the inequality~\eqref{eqn:ddr:shearer} has a rational witness $(\bm m, \bm s)$.
By multiplying the rational vectors $\bm \lambda$, $\bm w$, $\bm m$, and $\bm s$
with
the least common multiple of their denominators, we can convert
them to integer vectors.
Moreover, the vectors $\bm \lambda$, $\bm w$, $\bm m$, and $\bm s$ are non-negative.
We can represent a non-negative integer vector $\bm \lambda = (\lambda_{\bm Z})_{\bm Z \in\Sigma_{\out}}$ as a multiset $\calZ$ over $\Sigma_\out$
where for every $\bm Z \in\Sigma_{\out}$, the multiplicity of $\bm Z$ in $\calZ$ is equal to
$\lambda_{\bm Z}$. Similarly, we can represent $\bm w$, $\bm m$, and $\bm s$ as multisets $\calD$, $\calM$, and $\calS$ over $\Delta$, $\mon$, and $\sub$ respectively, leading to the following corollary.

\begin{corollary} \label{cor:integral:witness}
For rational coefficients $\bm\lambda$ and $\bm w$, inequality~\eqref{eqn:ddr:shearer} holds for all
polymatroids if and only if there exist multisets $\calZ$, $\calD$, $\calM$, and
$\calS$ over $\Sigma_{\out}$, $\Delta$, $\mon$, and $\sub$ respectively, such that the
following identity holds symbolically over the variables $h(\bm X)$, $\bm X \subseteq \bm
V$:
\begin{align}
  \sum_{\bm Z \in \calZ} h(\bm Z|\emptyset)
    &= \sum_{\delta \in \calD} h(\delta)
    - \sum_{\mu \in \calM} h(\mu)
    - \sum_{\sigma \in \calS} h(\sigma) \label{eq:identity:integers}
\end{align}

In particular, if we set $h(\emptyset) = 0$, then the identity~\eqref{eq:identity:integers} becomes:
\begin{align}
    \sum_{\bm Z \in \calZ} h(\bm Z)
      = \sum_{\delta \in \calD} h(\delta)
      - \sum_{\mu \in \calM} h(\mu)
      - \sum_{\sigma \in \calS} h(\sigma)
      \label{eqn:identity:no:emptyset}
\end{align}
\end{corollary}
The sizes of the multisets $\calZ, \calD, \calM, \calS$ are bounded by functions of
$\bm w$ and $|\bm V|$.

\begin{definition}
We call the terms $h(\delta)$ in~\eqref{eq:identity:integers} and~\eqref{eqn:identity:no:emptyset} {\em
statistics terms}, and call the terms $h(\mu)$ and
$h(\sigma)$ {\em witness terms}.
Specifically, we call terms $h(\mu)$ {\em monotonicity terms}, and terms $h(\sigma)$ {\em submodularity terms}.
After removing the common denominator in~\eqref{eqn:ddr:shearer}, the resulting inequality
is called an {\em integral Shannon inequality} and has the format:
\begin{align}
\sum_{\bm Z \in \calZ}h(\bm Z) \leq \sum_{\delta \in \calD} h(\delta)
\label{eqn:integral:ddr:shearer}
\end{align}
\end{definition}

\subsection{The Reset Lemma}
\label{subsec:reset}

To conclude this section on information inequalities, we present a combinatorial lemma that
plays a key role in our algorithm. The lemma says that, given a valid integral Shannon
inequality~\eqref{eqn:integral:ddr:shearer}, if we would like to remove an unconditional
term $h(\delta_{0})$ from the RHS while retaining its validity, then it suffices to remove
at most one term $h(\bm Z)$ from the LHS. We may be able to remove even more terms from the
RHS, in addition to $h(\delta_{0})$, but, importantly, it suffices to remove a single term
from the LHS.

\begin{lemma}[Reset Lemma] \label{lemma:reset} Consider an integral Shannon inequality~\eqref{eqn:integral:ddr:shearer}:
  \begin{align*}
    \sum_{\bm Z \in \calZ}h(\bm Z) \leq \sum_{\delta \in \calD} h(\delta)
  \end{align*}
  Suppose some term $\delta_0 \in \calD$ is {\em unconditional}, then there are two
  multisets $\calD' \subseteq \calD \setminus \{\delta_0\}$ and $\calZ' \subseteq \calZ$
  with $|\calZ'| \geq |\calZ|-1$
  such that the following is also an integral Shannon inequality:
  \begin{align*}
    \sum_{\bm Z \in \calZ'}h(\bm Z) \leq \sum_{\delta \in \calD'} h(\delta)
  \end{align*}
\end{lemma}

\begin{proof} By Corollary~\ref{cor:integral:witness}, there exists an integral witness such
that equation~\eqref{eqn:identity:no:emptyset} is an identity (with $h(\emptyset)$ set to $0$).
As mentioned before, this integral witness can be computed by solving a linear program
which gives a rational solution $(\bm m, \bm s)$, and then multiplying the rational vectors with the least common multiple of their denominators to convert them to integer vectors, which are then represented as the multisets $\calM$ and $\calS$ respectively.
We prove the lemma by induction on the ``potential'' quantity $q:=|\calD|+|\calM|+2|\calS|$.
The base case when $q \leq 1$ is trivial.
Consider the case when $q \geq 2$.
Suppose $\delta_0 = (\bm W \mid \emptyset)$, i.e.,~$h(\delta_0) = h(\bm W)$.

Since~\eqref{eqn:identity:no:emptyset} is an identity, the term $h(\bm W)$ must cancel out
with some other term. If $\bm W \in\calZ$, then we can remove $h(\bm W)$ from both sides
(i.e.,~setting $\calZ' = \calZ - \{\bm W\}$ and
$\calD' = \calD - \{(\bm W|\emptyset)\}$). Otherwise, there are three cases:

\begin{description}
  \item[Case 1]  There exists $(\bm Y|\bm W) \in \calD$.
    Then, from
    \begin{align}
      h(\bm Y|\bm W) + h(\bm W|\emptyset) = h(\bm Y\bm W|\emptyset)
      \label{eqn:D:step}
    \end{align}
    identity~\eqref{eqn:identity:no:emptyset} remains an identity if we
    add $(\bm Y\bm W|\emptyset)$ and remove
    $\{(\bm Y|\bm W), (\bm W|\emptyset)\}$ from $\calD$.
    The potential $q$ decreases by 1, and thus by induction,
    we can drop the newly added statistics term $h(\bm Y\bm W)$ from the RHS of~\eqref{eqn:integral:ddr:shearer} while dropping at most one term from the LHS.

  \item[Case 2] $h(\bm W)$ cancels with a monotonicity term $h(\mu)$ where
    $\mu = (\bm Y|\bm X)$.  In other words, $\bm W = \bm X\bm Y$, and
    the RHS of~\eqref{eqn:identity:no:emptyset} contains the terms:
    \begin{align}
        h(\bm W) - h(\bm Y|\bm X) &= h(\bm W) - h(\bm X\bm Y)+h(\bm X) =  h(\bm X)
        \label{eqn:M:step}
    \end{align}
    We add $(\bm X|\emptyset)$ to $\calD$, remove $(\bm W|\emptyset)$ from $\calD$, and remove
    $\mu = (\bm Y|\bm X)$ from $\calM$, thus decreasing the potential $q$ by 1.  Then, we proceed
    inductively to eliminate the newly added statistics term~$h(\bm X)$.

    \item[Case 3] $h(\bm W)$ cancels with a submodularity term $h(\sigma)$ where
    $\sigma = (\bm Y;\bm Z|\bm X)$ and $\bm W = \bm X \bm Y$.
    In particular, the RHS of~\eqref{eqn:identity:no:emptyset} contains the terms:
    \begin{align}
      h(\bm W) -h(\bm Y;\bm Z|\bm X)&=
      h(\bm W) -h(\bm X\bm Y) - h(\bm X\bm Z) + h(\bm X) + h(\bm X \bm Y \bm Z) \nonumber \\
      &= h(\bm X \bm Y \bm Z)  - h(\bm Z|\bm X)
      \label{eqn:S:step}
    \end{align}
    We add $(\bm X\bm Y\bm Z|\emptyset)$ to $\calD$, drop $(\bm W|\emptyset)$ from $\calD$,
    drop $\sigma = (\bm Y;\bm Z|\bm X)$ from $\calS$, and add a new monotonicity measure
    $(\bm Z|\bm X)$ to $\calM$. Overall, the potential $q$ decreases by $1$.
    The proof follows by induction where we can eliminate the newly added statistics term $h(\bm X\bm Y\bm Z)$ from the RHS.
  \end{description}
\end{proof}

We illustrate the reset lemma with the following simple examples:

\begin{example}
  Consider Shearer's inequality $2h(XYZ) \leq h(XY)+h(YZ)+h(XZ)$, which we write in the
  form~\eqref{eqn:integral:ddr:shearer} as:\footnote{To be consistent
  with~\eqref{eqn:integral:ddr:shearer}, we should write it as $h(XYZ)+h(XYZ)\leq
  h(XY|\emptyset)+h(YZ|\emptyset)+h(XZ|\emptyset)$.}
  \begin{align}
    h(XYZ)+h(XYZ) \leq & h(XY)+h(YZ)+h(XZ) \label{eq:triangle:reset}
  \end{align}
  To drop the term $h(XZ)$ on the RHS of~\eqref{eq:triangle:reset} while retaining the
  validity of the inequality, we can also delete one term on the LHS (we can choose any term since they are
  identical), and obtain the following Shannon inequality:
  \begin{align*}
    h(XYZ) & \leq h(XY)+h(YZ)
  \end{align*}
  The proof of the Reset Lemma ``explains'' how we can do this dropping by reasoning
  from the identity counterpart of~\eqref{eq:triangle:reset} (i.e.,~an identity of
the form~\eqref{eqn:identity:no:emptyset}):
  \begin{align}
    h(XYZ)+h(XYZ)
      &= h(XY)+h(YZ)+h(XZ)\nonumber\\
    &- (h(XY)+h(XZ)-h(X)-h(XYZ)) && \text{this is } h(Y;Z | X)\nonumber\\
    &-  (h(X)+h(YZ)-h(XYZ)-h(\emptyset)) && \text{this is } h(X; YZ | \emptyset)
  \label{eq:triangle:reset:identity}
  \end{align}
  By canceling out $h(XYZ)$ from both sides, we obtain a different identity:
  \begin{align}
    h(XYZ)&= h(XY)+h(YZ)\nonumber\\
&-(h(XY) - h(X)) && \text{this is a monotonicity } h(Y|X)\nonumber\\
&-(h(X)+h(YZ)-h(XYZ)-h(\emptyset))  && \text{this is } h(X; YZ | \emptyset)\label{eq:triangle:reset:identity2}
  \end{align}
which is what Case 3 from the proof of the Reset Lemma does.
The resulting identity witnesses the fact that $h(XYZ) \leq h(XY)+h(YZ)$ is a Shannon inequality.
\end{example}

\begin{example}
  Consider the following Shannon inequality:
  \[h(XYZW) + h(Y) \leq h(XY) + h(YZ) + h(W|XYZ),\]
  which follows from the following identity of the form~\eqref{eqn:identity:no:emptyset}:
  \[h(XYZW) + h(Y) = h(XY) + h(YZ) + h(W|XYZ) - h(X;Z|Y)\]
  Suppose that we
  want to drop $h(XY)$ from the RHS. The first step is going to
  follow Case 3 of the proof of the Reset Lemma by
  replacing the submodularity $h(X;Z|Y)$ with an additional monotonicity $h(Z|Y)$, thus leading to the identity:
  \[h(XYZW) + h(Y) = h(XYZ) + h(YZ) + h(W|XYZ) - h(Z|Y)\] where our target
  now is to remove the term $h(XYZ)$ from the RHS.  But now, our
  only option is to use Case 1 and combine $h(XYZ)$ with $h(W|XYZ)$, leading to
  \[h(XYZW) + h(Y) = h(XYZW) + h(YZ) - h(Z|Y)\] And now, we drop $h(XYZW)$
  from both sides. The resulting inequality is $h(Y) \leq h(YZ)$.
\end{example}

We will also need the following simple proposition to explain a step in \panda.

\begin{proposition}
Consider an integral Shannon inequality of the form~\eqref{eqn:integral:ddr:shearer}.
The number of unconditional statistics terms in $\calD$ (counting multiplicities)
is at least $|\calZ|$.
\label{prop:whole:terms}
\end{proposition}
\begin{proof}
It is straightforward to verify that the following vector $\bm h$ is a polymatroid,
i.e.,~satisfies~\eqref{eq:shannon}:
\begin{align*}
  h(\bm W) = &
               \begin{cases}
                 0 & \mbox{if $\bm W = \emptyset$}\\
                 1 & \mbox{otherwise}
               \end{cases}
\end{align*}
Applying $\bm h$ to~\eqref{eqn:identity:no:emptyset}, the LHS is $|\calZ|$, while the RHS
is $\leq$ to the number of unconditional terms, because
$h(\bm Y | \emptyset) =1$, and
$h(\bm Y | \bm X) =0$ when $\bm X \neq \emptyset$, while all witness terms are non-negative:
$h(\mu)\geq 0$, $h(\sigma)\geq 0$.
\end{proof}

\section{Overview of $\panda$ and statement of main result}
\label{sec:main:results}
This section states and explains the main result in this paper, which shows that the
Shannon inequality~\eqref{eqn:ddr:shearer} is not only useful for bounding the output size
of a DDR as shown in Theorem~\ref{th:upper:bound}, but it can also be used to guide an algorithm
to evaluate a DDR in time proportional to the bound~\eqref{eq:q:inequality}, modulo a
polylogarithmic factor.
We formally state this result in Section~\ref{subsec:main:result}, and present an illustrative
example in Section~\ref{subsec:example2}.
The detailed algorithm description and its proof of correctness are presented in Section~\ref{sec:algorithm}.

\subsection{An Efficient Algorithm to Evaluate Disjunctive Datalog Rules}
\label{subsec:main:result}

Given the coefficients $\bm w$ and $\bm \lambda$ of inequality~\eqref{eqn:ddr:shearer}
(where recall $\norm{\bm \lambda}_1 = 1$), we denote by:
\begin{align}
  B_{\Delta,\bm N} \defeq & \prod_{\delta \in \Delta} N^{w_\delta}_\delta
  & b_{\Delta,\bm n} \defeq
  & \log  B_{\Delta,\bm N} = \sum_{\delta \in \Delta} w_\delta n_\delta \label{eq:B:b}
\end{align}
Theorem~\ref{th:upper:bound} implies that, if $\Sigma_{\inn}$ satisfies the degree
constraints (see Sec~\ref{subsec:degrees}), i.e.,~$\Sigma_{\inn} \models (\Delta,\bm N)$,
then there exists a feasible output $\Sigma_{\out}$ that satisfies
$\norm{\Sigma_{\out}} \leq |\Sigma_\out|\cdot
B_{\Delta,\bm N}$.  Our main result states that such an output can be computed in time
$\tilde O(\norm{\Sigma_\inn}+B_{\Delta,\bm N})$ {\em if} inequality~\eqref{eqn:ddr:shearer}
is a Shannon inequality: (We use $\tilde O$ to hide a polylogarithmic factor in the input
size $\norm{\Sigma_{\inn}}$.)

\begin{theorem}[The \panda algorithm for a DDR] \label{th:algorithm}
    Given the following inputs:
    \begin{itemize}
        \item A disjunctive datalog rule (DDR) of the form~\eqref{eq:disjunctive:rule}.
        \item An input instance $\Sigma_\inn$ to the DDR.
        \item Degree constraints $(\Delta,\bm N)$ that are satisfied by the instance $\Sigma_\inn$,
            i.e.,~$\Sigma_\inn \models (\Delta, \bm N)$.
        \item Coefficients
          $\bm w = (w_\delta)_{\delta \in \Delta}$,
          and $\bm \lambda = (\lambda_{\bm Z})_{\bm Z \in \Sigma_\out}$, with $\norm{\bm\lambda}_1=1$
          that define a valid Shannon inequality~\eqref{eqn:ddr:shearer}.
    \end{itemize}
Let $B_{\Delta, \bm N}$ be the bound defined in~\eqref{eq:B:b}, in terms of the statistics
$(\Delta, \bm N)$ and the coefficients $\bm w$.
Then, we can compute a feasible output $\Sigma_{\out}$ to the
  DDR
  in time $\tilde O(\norm{\Sigma_\inn} + B_{\Delta,\bm N})$.
  This feasible output is of size $\tilde O(B_{\Delta,\bm N})$.
\end{theorem}

We will prove the theorem over the next few sections.  We illustrate with a simple example:

\begin{example} \label{ex:ab:query:continued} Continuing the example in
  Eq.~\eqref{eq:ab:query}, assume that $|R|=|S|=|U|=N$.  Consider the
  following Shannon inequality, proved by applying two submodularities:
  \begin{align*}
    h(XY)+h(YZ)+h(ZW) & \geq h(XYZ) + h(Y) + h(ZW) \geq h(XYZ)+h(YZW)
  \end{align*}
  Rearranging it to the form~\eqref{eqn:ddr:shearer}, we obtain:
  \begin{align}
    \frac{1}{2}h(XYZ) + \frac{1}{2}h(YZW) \leq \frac{1}{2}h(XY)+\frac{1}{2}h(YZ)+\frac{1}{2}h(ZW)
    \label{eq:panda:example:shannon}
  \end{align}
  Theorem~\ref{th:upper:bound} proves that there exists a feasible
  solution $(A,B)$ of size $|A|+|B| \leq 2N^{3/2}$, while
  Theorem~\ref{th:algorithm} says that we can compute a feasible
  solution of size $\tilde O(N^{3/2})$ in time $\tilde O(N^{3/2})$.
\end{example}

Theorem~\ref{th:algorithm} (and $\panda$) can be used to compute a full conjunctive query;
because a full conjunctive query is a special case of a DDR, where the output schema
consists of a single atom containing all variables. To that end, after computing a feasible output of the
DDR, we can add this output as an extra atom to the input schema and semijoin-reduce this atom with every other atom in order to obtain a solution to the
full conjunctive query.
Notice that because this extra atom contains all variables, adding it to the input schema
makes the query {\em acyclic}, hence semijoin reductions are now guaranteed to produce the correct output to the full conjunctive query with no extra tuples~\cite{DBLP:conf/vldb/Yannakakis81}.
Note also that the output of the full conjunctive query is a {\em minimal} model to the DDR
(Definition~\ref{def:feasible}).

\begin{corollary}[The $\panda$ algorithm for a full conjunctive query]
    \label{cor:full:cq}
  Suppose that the DDR from Theorem~\ref{th:algorithm}
  corresponds to a full conjunctive query $Q$, i.e.,~the output schema of the DDR consists of a single atom containing all variables $\Sigma_{\out}=\set{Q(\bm V)}$, in which case
  the DDR~\eqref{eq:disjunctive:rule} collapses back to~\eqref{eq:full:cq}, where $\Sigma := \Sigma_{\inn}$.
  Moreover, suppose that the conditions of Theorem~\ref{th:algorithm} are satisfied.
  Then, the output of the full conjunctive query $Q$ satisfies $|Q| \leq B_{\Delta,\bm N}$, and can be
  computed in time $\tilde O(\norm{\Sigma} + B_{\Delta,\bm N})$.
\end{corollary}

For the best-possible runtime, we want to seed $\panda$ with parameters minimizing the quantity $B_{\Delta,\bm N}$
(or equivalently, $b_{\Delta,\bm N}$),
which is the polymatroid bound~\eqref{eqn:polymatroid:bound} for the query $Q$.
In the case of full CQs, there is ony one $\bm \lambda$-parameter, set to $\lambda=1$, and it remains to find
$\bm w$ minimizing $B_{\Delta,\bm N}$. The minimum value is
$B^*_{\Delta, \bm N} \defeq 2^{b^*_{\Delta,\bm N}}$, defined by
\begin{align*}
b^*_{\Delta,\bm N} &\defeq \min_{\bm w \geq \bm 0} \left\{
  \sum_{\delta \in \Delta} w_\delta n_\delta
  \mid
  h(\bm V) \leq \sum_{\delta \in \Delta} w_\delta h(\delta) \text{ for all polymatroids } \bm h \in \R_+^{2^{\bm V}}
  \right\} \\
  &= \max_{\bm h \in \R_+^{2^{\bm V}}} \biggl\{
      h(\bm V) \mid h(\delta) \leq n_\delta \text{ for all } \delta \in \Delta
     \text{ and } \bm h \text{ is a polymatroid}
    \biggr\}
\end{align*}
The second equality follows from simple duality arguments.
$\panda$ would be worst-case optimal if
$B^*_{\Delta,\bm N}$ is tight, in the sense that we can construct a database instance
$\Sigma$ that satisfies the degree constraints $(\Delta,\bm N)$ and has output size $|Q| =
\Omega(B^*_{\Delta,\bm N})$. There are two special cases where $B^*_{\Delta,\bm N}$ is known
to be tight in data complexity: when all degree constraints are ``simple''
(see~\cite{DBLP:journals/tods/KhamisKNS21, 10175769}), or when the degree constraints are
``acyclic'' (see~\cite{DBLP:conf/pods/KhamisNS16,10.1145/3196959.3196990}). The degree
constraints are {\em simple} if $|\bm X| \leq 1$ for all $\delta = (\bm Y|\bm X) \in \Delta$, and
they are {\em acyclic} if there is a global order of the variables in $\bm V$ such that for every
$\delta = (\bm Y|\bm X) \in \Delta$, all variables in $\bm X$ precede all variables in $\bm
Y$ in the order.
If $\Delta$ contains only cardinality constraints, then it is both simple and acyclic,
and $B^*_{\Delta,\bm N}$ is exactly the \agm bound because~\eqref{eqn:ddr:shearer} reduces back to~\eqref{eqn:shearer}.

\subsection{Example: Preview of  \panda}
\label{subsec:example2}

We illustrate the algorithm with the DDR~\eqref{eq:ab:query}:
\begin{align*}
  A(X,Y,Z) \vee B(Y,Z,W) \cd & R(X,Y) \wedge S(Y,Z) \wedge U(Z,W)
\end{align*}
Following Example~\ref{ex:ab:query:continued}, we assume we only have cardinality statistics / constraints:
\begin{align*}
  \Delta = & \set{(XY|\emptyset), (YZ|\emptyset), (ZW|\emptyset)}, & N_{XY}=&N_{YZ}=N_{ZW}=N.
\end{align*}
We further assume that $N$ is a power of 2.
Each constraint has a ``guard'', i.e.,~a relation that ``sponsors'' (satisfies)
the constraint:
$T_{XY}:=R, T_{YZ}:=S, T_{ZW}:=U$. We name all guards $T_{\cdot}$ to be consistent with the
formal description of the algorithm presented in Section~\ref{sec:algorithm}.

Suppose that the input Shannon inequality given is the one in Eq~\eqref{eq:panda:example:shannon}.
This inequality is in the shape of~\eqref{eqn:ddr:shearer} with
$\lambda_{XYZ}=\lambda_{YZW}=1/2$ and
$w_{XY|\emptyset}=w_{YZ|\emptyset}=w_{ZW|\emptyset}=1/2$. The fact that this is a Shannon
inequality was shown above in Example~\ref{ex:ab:query:continued}.

From Corollary~\ref{cor:integral:witness}, the Shannon inequality~\eqref{eq:panda:example:shannon}
must have a corresponding identity~\eqref{eqn:identity:no:emptyset} over symbolic
variables $h(\bm X)$ where $h(\emptyset) = 0$. We show one such identity below:\footnote{Our witness is not elemental, because $h(Y;ZW|\emptyset)$ is not elemental:
if we replaced the latter with $h(Y;W|\emptyset)+h(Y;Z|W)$, then we obtain an elemental
witness.  Also note that the witness is not unique; for example, we could have used the
witness $h(XY;Z|\emptyset) + h(Y;W|Z)$.}
\begin{align}
  h(XYZ)+h(YZW) &= h(XY|\emptyset)+h(YZ|\emptyset)+h(ZW|\emptyset)
    - h(X;Z|Y) - h(Y;ZW|\emptyset) \label{eq:ex:identity:integers}
\end{align}
The bound from~\eqref{eq:B:b} is
$b_{\Delta, \bm N} = (n_{XY}+n_{YZ}+n_{ZW})/2 = 3n/2$ where $n \defeq \log N$, or
equivalently, $B_{\Delta, \bm N} = N^{3/2}$; this is the runtime budget the algorithm cannot
exceed, in order to compute a feasible solution $(A, B)$ to the DDR.

\begin{figure}[th!]
  \centering
  \includegraphics[width=\textwidth]{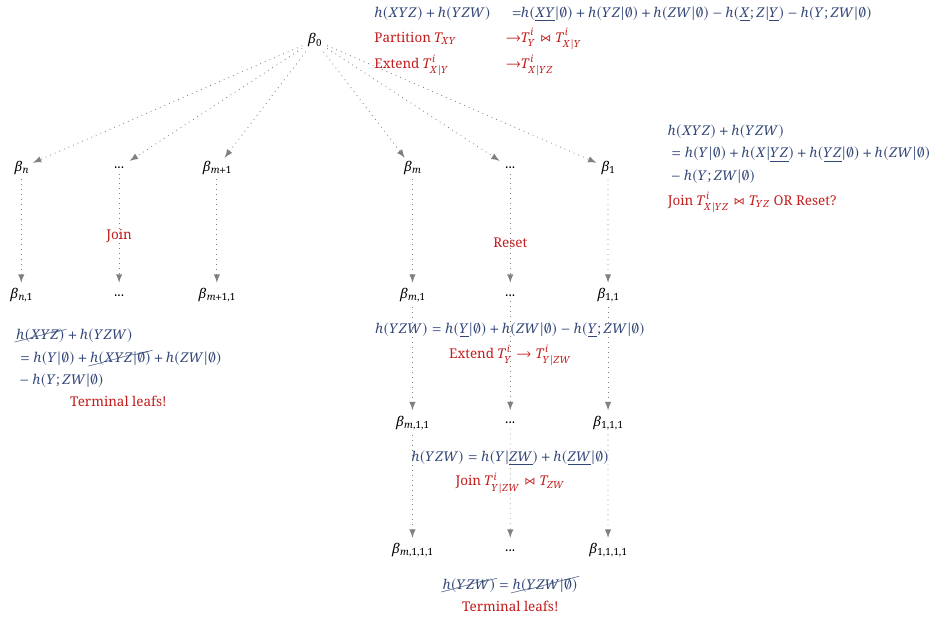}

\caption{An example illustrating the \panda algorithm over the disjunctive rule~\eqref{eq:ab:query}. Here, $n \defeq \log N$ and $m \defeq \frac{n}{2}$.
For each node in the sub-problem tree,
the corresponding
identity~\eqref{eqn:identity:no:emptyset} is in {\color{@ThCSdarkblue}blue} while the corresponding algorithmic operation is in {\color{@ThCSdarkred}red}.}
\label{fig:example}
\end{figure}

Our algorithm operates by observing and
manipulating identity~\eqref{eq:ex:identity:integers}, while maintaining its identity
invariant. Every step of the algorithm applies a modification to the identity and mirrors
this modification with an algorithmic computation to create
one or more sub-problems. Each sub-problem is then equipped with the newly created identity,
the newly created input data, and carry on the next step on their own.
The spawned sub-problems form a (sub-problem) tree. In the end, we gather results from
the leaves of this tree to answer the original query.

In identity~\eqref{eq:ex:identity:integers}, the statistics terms
$h(XY|\emptyset)$, $h(YZ|\emptyset)$, and $h(ZW|\emptyset)$ correspond to input data
(input relations) $R$, $S$, and $U$.
When we modify the identity, we will be transforming some subsets of these statistics
terms into different statistics terms; correspondingly, the input relations are
manipulated to create new input relations that are associated with the new statistics
terms.
We next describe this idea in more details.
The steps of the algorithm are also shown in Figure~\ref{fig:example}.

In the first step, we grab an unconditional statistics term (from $\calD$)
in~\eqref{eq:ex:identity:integers}. Let's say the term we grab is $h(XY |\emptyset)=h(XY)$.
Since~\eqref{eq:ex:identity:integers} is an identity, there must be another term that
cancels out the symbolic variable $h(XY)$. In this case, it is $-h(X;Z|Y)$; so we combines
$h(XY)$ with the canceling term to obtain new statistics terms:
\begin{align*}
  h(\underline{XY}|\emptyset) - h(\underline{X};Z|\underline{Y}) &=
  h(XY)-h(\emptyset)-h(XY)-h(YZ)+h(XYZ)+h(Y)\\
  &=h(Y) - h(\emptyset)+ h(XYZ)-h(YZ)\\
  &=h(Y|\emptyset)+h(X|YZ)
\end{align*}
This rewriting-by-cancellation changed our identity~\eqref{eq:ex:identity:integers} into a new one:
\begin{align} \label{eqn:after:step:1}
  h(XYZ)+h(YZW) &= h(Y|\emptyset)+h(X|YZ)+h(YZ|\emptyset)+h(ZW|\emptyset) - h(Y;ZW|\emptyset)
\end{align}
The change amounts to replacing a statistics term with two new ones:
$h(XY|\emptyset) \to h(Y|\emptyset)+h(X|YZ)$.
We mimic this symbolic change with an actual relational change that shall alter the
structure of the problem.
The rough idea is to create inputs to the new problem by creating two new guards for the two new
statistics terms:
\begin{itemize}
\item The guard for $h(Y|\emptyset)$ will be the table $T_Y \defeq \pi_Y(R)$.
\item The guard for $h(X|YZ)$ is obtained by using $R(X,Y)$ to construct a ``dictionary''
(a lookup table) which supports the following operation: given
a tuple $(y, z)$, return the list of all $x$-values such that $(x, y) \in R$.
(Note that this operation does not use the given $z$-value.)
We denote this dictionary as $T_{X|YZ}$.
\end{itemize}
The aim is for us to be able to join all $4$ guards (corresponding to the four statistics
terms on the RHS of~\eqref{eqn:after:step:1}) to answer the same query as before.
However, implementing this idea straight up does not work, because the new degree constraint
for $T_Y$ is $N_Y \defeq |\pi_Y(R)|$ and for $T_{X|YZ}$ is $N_{X|YZ} \defeq\deg_R(X|Y)$, and they are too
large: the product $N_Y \cdot N_{X|YZ}$ can be much greater than the original statistics
of $N_{XY} = N = |R|$ (guarding $h(XY)$ that we started with).
If this product is too large, we cannot apply induction to solve the sub-problem in our time budget of $B_{\Delta, \bm N} = N^{3/2}$.

The reason this product $N_Y\cdot N_{X|YZ}$ is too large is that $N_Y$ counts both
high-degree and low-degree $y$-values, while the new statistics $N_{X|YZ}$ is the maximum
degree. Thus, the product is an over-estimation of what the size of the join $T_Y \Join
T_{X|YZ}$ can be. To remedy the situation, we {\em uniformize} the problem by partitioning
$R$ into a logarithmic number of sub-relations $R = R^1 \cup \cdots \cup R^k$, where each
sub-relation contains tuples whose $y$-values have similar degrees. In effect,
uniformization is an algorithmic technique for dealing with {\em skews}, a notoriously
well-known reason for why query plans might blow up in practice.\footnote{To
briefly explain  this insight, consider the triangle query $Q(x,y,z) \cd R(x,y) \wedge
S(y,z) \wedge T(x,z)$. It is straightforward to construct input instances for which $|R|=|S|=|T|=N$,
and $|R \Join S|$, $|S \Join T|$, and $|R \Join T|$ are all $\Omega(N^2)$, by having high-degree
(i.e.,~skewed) values to join~\cite{DBLP:journals/jacm/NgoPRR18}.
To overcome this limitation of join-project query plans, one variant of worst-case optimal
join algorithms does the following. Partition $R = R^h \cup R^\ell$, where
$R^h \defeq \set{(x,y) \in R \mid \degree_R(Y|X=x) > \sqrt N}$ and
$R^\ell \defeq R \setminus R^h$.\newline
The query can be rewritten as $Q(x,y,z) \cd R^h(x,y) \wedge S(y,z) \wedge T(x,z) \vee
R^\ell(x,y) \wedge S(y,z) \wedge T(x,z)$. The joins $R^h(x,y) \wedge S(y,z)$ and $R^\ell(x,
y) \wedge T(x, z)$ are both cardinality-bounded by $N^{3/2}$ and computable within
$O(N^{3/2})$-time. Thus, the entire query can be computed within $O(N^{3/2})$-time.\newline
Our uniformization step is a generalization of this idea to the setting of DDRs. We needed
to partition $R$ into $k$ parts instead of $2$ parts to maintain our algorithmic invariants.
For some special classes of queries, Bringmann and
Gorbachev~\cite{bringmann2024} show that
heavy/light partitioning is sufficient to achieve the same result.}

Concretely, the partitioning is done as follows.
Relation $R^i$ contains all tuples $(x,y)\in R$ where:
\begin{align*}
  \frac{N}{2^i}\leq & \degree_{R}(X|Y=y) < \frac{N}{2^{i-1}}
\end{align*}
We set $T^i_Y \defeq \pi_Y(R^i)$, thus $|T_Y^i| \leq 2^i$.  We further
partition $T_Y^i$ into two equal-sized buckets (which we will continue
to name ``buckets $i$'', with some abuse).  The two new guards $T_Y^i$ and $R^i$ have
statistics $N^i_Y$ and $N^i_{X|YZ}$, which (by our partition-into-two
trick) satisfy the following:
\begin{align*}
    N^i_Y \defeq |T_Y^i| & \leq 2^{i-1} \\
    N^i_{X|YZ} \defeq\degree_{R^i}(X|Y) &\leq \frac{N}{2^{i-1}} \\
   N^i_Y \cdot N^i_{X|YZ} & \leq N
\end{align*}
To be consistent with the notation used in describing \panda,
all guards will be called $T$: in particular, the two new guards  $\pi_Y(R^i)$ and $R^i$ are called $T^i_Y$ and
$T^i_{X|YZ}$ respectively.

After partitioning, we now have $O(\log N)$ sub-problems, each equipped with
identity~\eqref{eqn:after:step:1}. In the second step, we again grab an unconditional
statistics term $h(YZ|\emptyset)$, and find a term from~\eqref{eqn:after:step:1} to
cancel it.\footnote{Note how ``greedy'' the algorithm is. This is one of the reasons for the
large polylog factor it suffers.} This time, the cancellation is:
\begin{align*}
  h(X|\underline{YZ})+h(\underline{YZ}|\emptyset) &=
  h(XYZ)-h(YZ)+h(YZ)-h(\emptyset) =h(XYZ|\emptyset),
\end{align*}
leading to a new identity
\begin{align} \label{eqn:after:step:2}
  h(XYZ)+h(YZW) &= h(Y|\emptyset)+h(XYZ|\emptyset) + h(ZW|\emptyset)- h(Y;ZW|\emptyset)
\end{align}
The change $h(X|\underline{YZ})+h(\underline{YZ}|\emptyset) \to h(XYZ)$ is suggestive of a
join, where for the $i$th subproblem, we will join the corresponding guards : $T^i_{X|YZ}
\Join T_{YZ}$. Recall that $N^i_{X|YZ} \leq N/2^{i-1}$, hence for $i > \frac 1 2 \log N$
performing this join won't take time over the budget of $O(N^{3/2})$. On the other hand,
when $i \leq \frac 1 2 \log N$, we will need an additional idea, to use the reset lemma
(Lemma~\ref{lemma:reset}) to reroute the sub-problem away from the ``heavy-hitter hotspot'',~i.e.,~join over the high-degree $Y$-values. In
addition to uniformization, the reset lemma is our second ingredient to deal with skews.

Concretely, for the $i$th sub-problem, we do the following:
\begin{itemize}
\item When $i > \frac 1 2 \log N$, then \panda performs a join
  $T^i_{XYZ} := T^i_{X|YZ} \Join T_{YZ}$.
  The output size of this join is within the bound $B_{\Delta, \bm N} =
  N^{3/2}$. After computing this join, there will be no more sub-problems because
  we have computed a relation that fits the output schema,
  namely it corresponds to $A(X, Y, Z)$.
  We add tuples from $T^i_{XYZ}$ to the output relation $A(X,Y,Z)$.

\item In the case where $i \leq \frac 1 2 \log N$, the algorithm attempts to
  perform the same join $T^i_{XYZ} := T^i_{X|YZ} \Join T_{YZ}$, but its
  output size now exceeds the bound $B_{\Delta, \bm N} = N^{3/2}$.  Therefore, the
  algorithm does not compute a guard $T^i_{XYZ}$ for $h(XYZ|\emptyset)$, but
  instead uses the reset lemma to cancel out this term with the term
  $h(XYZ|\emptyset)$ on the LHS.
  The new identity (for these sub-problems) is now
    \begin{align*}
      h(YZW) &= h(Y|\emptyset) + h(ZW|\emptyset)- h(Y;ZW|\emptyset)
    \end{align*}
  We again grab an unconditional statistics term $h(Y|\emptyset)$
  and cancel it with $h(\underline{Y};ZW|\emptyset)$:
    \begin{align*}
      h(Y|\emptyset)-h(Y;ZW|\emptyset) &= h(Y|ZW)
    \end{align*}
    The guard $T^i_Y$ for $h(Y|\emptyset)$ has size $|T^i_Y| \leq N^{1/2}$,
    thanks to the fact that $|T^i_Y| \leq 2^{i-1}$ and $i \leq \frac{1}{2}\log N$,
  The above step replaces $h(Y|\emptyset)$ with a new statistics term $h(Y|ZW)$.  Its guard is computed from $T^i_Y$, by extending it
  into a dictionary $T^i_{Y|ZW}$: given $(z,w)$, this dictionary returns all $y$-values for which $(y) \in T^i_Y$.
  In particular, this silly dictionary always returns the entire table $T^i_Y$, no matter what the given $(z, w)$ are.\footnote{
    To be more precise, before \panda extends $T^i_Y$ into $T^i_{Y|ZW}$, it will ``uniformize'' $T^i_Y$ by partitioning the table $T^i_Y$ into $\log |T^i_Y|$
  buckets based on the ``degree'' $\degree_{T^i_Y}(Y|\emptyset)$.
  However, this partition is vacuous since only one bucket will be non-empty.}
  After the dictionary extension, the algorithm performs a join
  $T^{i}_{YZW}:=T^i_{Y|ZW}\Join T_{ZW}$.
  This join is feasible since the output size is
  within the bound $B_{\Delta, \bm N} = N^{3/2}$.  Now, the algorithm reaches a
  terminal node since the join result $T^{i}_{YZW}$ is in the output
  schema; namely, it corresponds to $B(Y, Z, W)$.
\end{itemize}

At the end of the algorithm, the tables $T^i_{XYZ}$ from all branches are unioned into the
output relation $A(X, Y, Z)$, while tables $T^i_{YZW}$ from all branches are unioned into the
other output relation $B(Y, Z, W)$. These two relations are the final output of the
algorithm for the DDR~\eqref{eq:ab:query}.
Note that the input table $T_{ZW}$ does not contribute to the output table $A(X,Y,Z)$, and similarly the input table $T_{YZ}$ does not contribute to the output table $B(Y,Z,W)$.
Nevertheless, the tables $A(X,Y,Z)$ and $B(Y,Z,W)$ are still a valid output to the DDR. This is because Definition~\ref{def:feasible} only requires that if a tuple $(x, y, z, w)$ satisfies the conjunction $R(x,y) \wedge S(y,z) \wedge U(z,w)$, then it must satisfy the disjunction $A(x,y,z) \vee B(y,z,w)$.
We do not require the converse to hold in DDR semantics. (The converse is only required in CQ semantics, as we will see in Section~\ref{sec:subw}.)

Note that, for the particular query~\eqref{eq:ab:query}, there is a way to compute a
feasible output without the polylog factor as described above. The above strategy is only
meant to illustrate the \panda algorithm in its full generality.

\section{Detailed Description of \panda}
\label{sec:algorithm}

This section describes the \panda algorithm in detail.
Section~\ref{subsec:tables:dictionaries} presents the main data structures (called tables
and dictionaries) used by the algorithm. Section~\ref{subsec:algorithm} presents the
algorithm and its proof of correctness and runtime analysis.

\subsection{Tables and Dictionaries}
\label{subsec:tables:dictionaries}

For a given statistics term $\delta = (\bm Y | \bm X)$, \panda uses two kinds of data
structures: {\em tables} and {\em dictionaries}, denoted $T_\delta$.  When $\bm
X=\emptyset$, then we call it a {\em table}; otherwise, we call it a {\em dictionary}.
There will be at most one table/dictionary $T_\delta$ for a given $\delta$. As usual, we
abbreviate $\bm Y|\emptyset$ with just $\bm Y$, and statistics
$N_{\bm Y|\emptyset}, n_{\bm Y|\emptyset}$
with just $N_{\bm Y}$ and $n_{\bm Y}$.
Specifically, a table is a set $T_{\bm Y} \subseteq \dom^{\bm Y}$ of tuples over the $\bm Y$ variables,
and a dictionary is a function $T_{\bm Y|\bm X} : \dom^{\bm X} \rightarrow 2^{\dom^{\bm Y}}$
that gives a set of tuples over the $\bm Y$ variables {\em given} a specific binding
$\bm X =\bm x$ of the $\bm X$-variables.

For a statistics term $\delta = (\bm Y | \bm X)$, each table/dictionary $T_\delta$ is
associated with a {\em statistics} $N_\delta$. We say that $T_\delta$ {\em satisfies} the
statistics, and we write $T_\delta \models N_\delta$, iff $|T_\delta(\bm x)| \leq N_\delta$
for all $\bm x \in \dom^{\bm X}$. As a special case, a table $T_{\bm Y}$ {\em satisfies} a
statistics $N_{\bm Y}$ iff $|T_{\bm Y}| \leq N_{\bm Y}$.

The algorithm performs the following operations on tables and dictionaries:
join, projection, extension, construction, and partition.
Each operation yields the corresponding statistics on the results, as described below.

\begin{itemize}
\item {\em Join} of a table with a dictionary,
  $T_{\bm X\bm Y} := T_{\bm X} \Join T_{\bm Y|\bm X}$.  This operation
  constructs a new table with statistics
  $N_{\bm X\bm Y} \defeq N_{\bm X} N_{\bm Y|\bm X}$.  The join takes time $O(N_{\bm X} N_{\bm Y|\bm X})$.
\item {\em Projection} of a table, $T_{\bm X} := \pi_{\bm X}(T_{\bm X\bm Y})$.
This operation takes a table $T_{\bm X \bm Y}$ over variables $\bm X \cup \bm Y$,
with statistics $N_{\bm X\bm Y}$, and constructs a new table $T_{\bm X}$ with
  statistics $N_{\bm X} \defeq N_{\bm X\bm Y}$. The projection takes time
  $O(N_{\bm X\bm Y})$.
\item {\em Extension} of a dictionary $T_{\bm Y|\bm X}$ into another dictionary $T_{\bm
Y|\bm X\bm Z}$, where $\bm Z$ is disjoint from $\bm X\bm Y$. This operation takes as input a
dictionary $T_{\bm Y|\bm X}$ and returns a new dictionary $T_{\bm Y|\bm X\bm Z}$ defined as
$T_{\bm Y|\bm X\bm Z}(\bm x,\bm z) \defeq T_{\bm Y|\bm X}(\bm x)$ for all $
(\bm x,\bm z) \in \dom^{\bm X\bm Z}$.  Its statistics is $N_{\bm Y|\bm X\bm Z} \defeq N_{\bm Y|\bm X}$.
This operation takes $O(1)$ time, because the operation does not touch the data, but
only constructs a new function that calls the existing function $T_{\bm Y|\bm X}$.
\item {\em Construction}.  Given a table $T_{\bm X\bm Y}$
  over variables $\bm X \cup \bm Y$ with statistics $N_{\bm X\bm Y}$, construct a dictionary
  $T_{\bm Y|\bm X}$, with statistics
  $N_{\bm Y|\bm X} \defeq \degree_{T_{\bm X\bm Y}}(\bm Y|\bm X)$.
  This operation takes $O(N_{\bm X\bm Y})$ time because it involves
  scanning the table $T_{\bm X\bm Y}$ and constructing an index.
  The operation returns a function that looks up in this index.
\item {\em Partition}. Given a table $T_{\bm X\bm Y}$
over variables $\bm X \cup \bm Y$ with statistics $N_{\bm X\bm Y}$,
partition $T_{\bm X\bm Y}$ into $k:=2\lceil \log |T_{\bm X\bm Y}| \rceil$ many sub-tables
$T^{1},\dots,T^{k}$ satisfying the conditions stated in Lemma~\ref{lmm:partition}
below. This operation takes time $O(N_{\bm X\bm Y})$.
\end{itemize}

\begin{lemma}
Let $T_{\bm X\bm Y}$ be a table over variables $\bm X \cup \bm Y$ with statistics $N_{\bm X\bm Y}$.
Then, $T_{\bm X\bm Y}$ can be partitioned into at most
$k := 2 \lceil \log |T_{\bm X\bm Y}| \rceil $ sub-tables $T^{1},\dots,T^{k}$
satisfying
\begin{align}
  |\pi_{\bm X}(T^{i})| \cdot \deg_{T^{i}}(\bm Y |\bm X) \leq N_{\bm X\bm Y}, \quad\quad \forall i \in [k].
  \label{eq:partition}
\end{align}
\label{lmm:partition}
\end{lemma}
\begin{proof}
To obtain the sub-tables $T^{i}$, observe that the number of tuples $\bm x \in
\pi_{\bm X}(T_{\bm X\bm Y})$ with $\log$-degree in the interval
$[i,i+1)$ is at most $|T_{\bm X\bm Y}|/2^i \leq 2^{n_{\bm X\bm Y}-i}$.
Hence, if we partition $T_{\bm X\bm Y}$ based on which of the buckets $[i,i+1)$ the
$\log$-degree falls into, we will almost attain the required inequality, just off by a factor of $2$:
\[ |\pi_{\bm X}(T^{i})| \cdot \deg_{T^{i}}(\bm Y |\bm X)  \leq 2^{(n_{\bm X\bm Y}-i)}\cdot2^{(i+1)} = 2N_{\bm X\bm Y}. \]
To get rid of the factor of $2$, we partition each $T^{i}$ into two
tables whose projections onto $\bm X$ are equally sized. Overall, we need
$k := 2\lceil \log |T_{\bm X\bm Y}| \rceil$ partitions.
\end{proof}

\subsection{Algorithm}
\label{subsec:algorithm}

This section describes \panda and proves Theorem~\ref{th:algorithm}.
The main input to \panda contains an input instance $\Sigma_{\inn}$,
an output schema $\Sigma_{\out}\neq \emptyset$, and
statistics $(\Delta, \bm N)$ satisfied by the input instance, i.e.,~$\Sigma_{\inn} \models (\Delta, \bm N)$
as defined in Section~\ref{subsec:degrees}.
In addition, we are also given the coefficients $\bm\lambda$ and $\bm w$
for which~\eqref{eqn:ddr:shearer} is a Shannon inequality.
By Proposition~\ref{prop:rational:shannon}, we can assume that $\bm\lambda$ and $\bm w$ are rational and independent of the statistics $(\Delta, \bm N)$.
From Corollary~\ref{cor:integral:witness}, we can also assume that \panda was given the multisets
$\calZ, \calD, \calM, \calS$ for which identity~\eqref{eqn:identity:no:emptyset} holds.
In particular, as in the proof of Corollary~\ref{cor:integral:witness},
given vectors $\bm \lambda$ and $\bm w$ that make Eq.~\eqref{eqn:ddr:shearer} a Shannon inequality,
the witness $(\bm m, \bm s)$ can be computed by solving a linear program.
Since $\bm \lambda$ and $\bm w$ are rational and independent of the data,
the solution $(\bm m, \bm s)$ is also rational and independent of the data.
By multiplying the vectors $\bm \lambda$, $\bm w$, $\bm m$, and $\bm s$ with the least common multiple of their denominators, we can convert them to integer vectors.
And now we can represent the integral $\bm\lambda, \bm w, \bm m$ and $\bm s$ as multisets $\calZ$, $\calD$, $\calM$, and $\calS$ respectively.
All the above steps take constant time in {\em data complexity}; see Section~\ref{subsec:conjunctive}.
This is because the vectors $\bm \lambda$, $\bm w$, $\bm m$, and $\bm s$ have dimension $O(1)$
in data complexity. Moreover, since the values of these vectors don't depend on the data, they are considered constants in data complexity.

We shall show that a feasible output $\Sigma_{\out}$ to the DDR~\eqref{eq:disjunctive:rule}
can be computed in time $\tilde O(\norm{\Sigma_\inn}+B_{\Delta,\bm N})$, defined in~\eqref{eq:B:b}. In terms of
these multisets, the bounds defined in~\eqref{eq:B:b} have the following equivalent
expressions:
\begin{align}
  B_{\Delta,\bm N}
  &= \left(\prod_{\delta \in \calD} N_{\delta}\right)^{1/|\calZ|}& b_{\Delta,\bm n} = \log B_{\Delta, \bm N}
  &= \frac{1}{|\calZ|} \sum_{\delta \in \calD} n_{\delta} \label{eq:B:b:integers}
\end{align}

For each statistics term $\delta = (\bm Y|\bm X) \in \calD$, there is a guard which is a
table/dictionary instance $T_{\bm Y|\bm X}$ with statistics $N_{\bm Y|\bm X}$, i.e.,~$T_{\bm
Y|\bm X} \models N_{\bm Y|\bm X}$ as defined in Section~\ref{subsec:tables:dictionaries}.
Creating the initial guards can be done in time $\tilde O(\norm{\Sigma_\inn})$.
We adopt the convention that the lower case letters
$n_\delta, b_{\Delta,\bm n}$ represent the logarithms of the upper case $N_\delta,
B_{\Delta,\bm N}$ respectively.

Summarized in Algorithm~\ref{algo:panda}, \panda works as follows.
Starting from an initial node, it grows a tree of sub-problems. New sub-problems are
spawned from a ``non-terminal'' leaf node of the tree. The process stops when every leaf
node is terminal, a concept we shall define shortly. After the tree growing stops,
a feasible output of the problem is gathered from the leaves of the tree.

\setlength{\algomargin}{2em}
\SetKwComment{Comment}{$\triangleright$\ }{}
\SetInd{0.5em}{0.5em}
\afterpage{\begin{algorithm}
    \caption{\panda}
    \label{algo:panda}
     Initialize the sub-problem tree with a single node with input parameters
               $(\calZ, \calD, \calM, \calS, \bm T, \bm n)$\;
          \While(\Comment*[f]{Tree-growing Loop}){$\exists$ a non-terminal leaf $\beta$}{
               Let $(\calZ, \calD, \calM, \calS, \bm T, \bm n)$
              be the parameters of this leaf\;
               Pick $\delta = (\bm W | \emptyset) \in \calD$ arbitrarily \label{line:arbitrary:delta} \Comment*[r]{See Prop~\ref{prop:correctness} for why such $\delta$ exists}
               \If(\Comment*[f]{Case 1 of Lemma~\ref{lemma:reset}: apply~\eqref{eqn:D:step}}){$\exists$ $(\bm Y| \bm W)\in \calD$\label{line:case1}}
               {\If{$n_{\bm W}+n_{\bm Y|\bm W} \leq b_{\Delta,\bm n}$\label{line:small:join}}{
                      $\bm n' = \bm n \cup \{n_{\bm Y\bm W}\} -
                                  \{ n_{\bm W} + n_{\bm Y|\bm W} \}$, where
                        $n_{\bm Y\bm W} := n_{\bm W} + n_{\bm Y|\bm W}$\;
                        $\calD' = \calD \cup \{ (\bm Y\bm W | \emptyset) \}  -
                                  \{ (\bm W|\emptyset), (\bm Y | \bm W)\}$\label{line:new:D'}\;
                        $\bm T' = \bm T \cup \{ T_{\bm Y\bm W} \} -
                             \{ T_{\bm W}, T_{\bm Y|\bm W}\}$,
                        where $T_{\bm Y\bm W} := T_{\bm W} \Join T_{\bm Y|\bm W}$, which guards $(\bm Y\bm W|\emptyset)$\;
                         Create a child of $\beta$, with parameters
                        $(\calZ, \calD', \calM, \calS, \bm T', \bm n')$\;
                        }
                   \Else{\label{line:else}
     $\overline \calD = \calD \cup \{ (\bm Y\bm W | \emptyset) \}  -
                                  \{ (\bm W|\emptyset), (\bm Y | \bm W)\}$\label{line:overline:calD}\;
                      Apply Lemma~\ref{lemma:reset} to $(\calZ, \overline \calD, \calM,\calS)$
                        to obtain $(\calZ', \calD', \calM',\calS')$
                        where $\calD' \subseteq \overline\calD - \{(\bm Y\bm W|\emptyset)\}$
                        \label{line:else:reset}\;
 Let $\bm T', \bm n'$ be $\bm T, \bm n$ with only the terms
                        corresponding to $\calD'$ retained\;
 Create a child of $\beta$, with parameters
                             $(\calZ', \calD', \calM', \calS', \bm T', \bm n')$
                             \label{line:calZ:2}\;
                           }
                         }
               \ElseIf(\Comment*[f]{Case 2 of Lemma~\ref{lemma:reset}: apply~\eqref{eqn:M:step}}){$\exists (\bm Y | \bm X) \in \calM$ with $\bm W=\bm X\bm Y$\label{line:case2}}{
                 $\bm n' = \bm n \cup \{ n_{\bm X} \} - \{ n_{\bm W}\}$
                        where $n_{\bm X} := n_{\bm W}$\;
                    $\calM' = \calM - \{(\bm Y|\bm X)\}$\;
                    $\calD' = \calD \cup \{ (\bm X| \emptyset) \} - \{(\bm W| \emptyset)\}$\;
                    $\bm T' = \bm T \cup \{T_{\bm X} \} - \{ T_{\bm W}\}$
                        where $T_{\bm X} := \pi_{\bm X}(T_{\bm W})$, which guards
                        $(\bm X | \emptyset)$\;
                    Create a child of $\beta$, with parameters
                   $(\calZ, \calD', \calM', \calS, \bm T', \bm n')$\;
                   }
              \ElseIf(\Comment*[f]{Case 3 of Lemma~\ref{lemma:reset}: apply~\eqref{eqn:S:step:algo} instead of~\eqref{eqn:S:step}}) {$\exists (\bm Y; \bm Z |\bm X) \in \calS$ with $\bm W=\bm X\bm Y$\label{line:case3}}{
                   Partition $T_{\bm W} = T^1 \cup T^2 \cup \cdots \cup T^k$ with $k := O(\log |T_{\bm W}|)$,
                   using Lemma~\ref{lmm:partition}\;
                   \For {$i \leftarrow 1$ to $k$\label{line:sub:problem}}{
                         $\bm n^i = \bm n \cup \{ n^i_{\bm X}, n^i_{\bm Y|\bm X\bm Z} \} -
                        \{ n_{\bm W}\}$
                        where $n^i_{\bm X} := \log |\pi_{\bm X}(T^i)|$ and $n^i_{\bm Y| \bm X\bm Z} := \log \deg_{T^i}(\bm Y|\bm X)$\label{line:sub:problem:stats}\;
                         $\calD^i = \calD \cup \{(\bm X|\emptyset), (\bm Y|\bm X\bm Z)\} - \{(\bm W|\emptyset)\}$\;
                        $\calS^i = \calS - \{(\bm Y; \bm Z|\bm X)\}$\;
                         $\bm T^i = \bm T \cup \{T^i_{\bm X}, T^i_{\bm Y | \bm X\bm Z}\} -
                             \{ T_{\bm W}\}$,
                        where $T^i_{\bm X} := \pi_{\bm X} (T^i)$ and
                        $T^i_{\bm Y | \bm X\bm Z}$ is an extension of
                        $T^i_{\bm Y | \bm X}$\;
                         Create the $i$th-child of $\beta$, with parameters
                             $(\calZ, \calD^i, \calM, \calS^i, \bm T^i, \bm n^i)$\;
                   }
              }
            }
            \;
   $Q(\bm Z) \leftarrow \emptyset$ for all $Q(\bm Z) \in \Sigma_{\out}$ \label{line:gathering}\;
          \For(\Comment*[f]{Result-gathering phase}){each terminal leaf $\beta$}{
             Let $(\calZ, \calD, \calM, \calS, \bm T, \bm n)$ be the parameters of this leaf\;
             Pick $\delta = (\bm Z | \emptyset) \in \calD$ such that
                   $\bm Z \in \Sigma_{\out}$ \Comment*[r]{By Definition~\ref{defn:terminal:leaf}}
             $Q(\bm Z) \leftarrow Q(\bm Z) \cup T_{\bm Z}$ \label{line:got:it}\;
                 }
\end{algorithm}
}

To every node, there associates a sub-problem parameterized by
a tuple $(\calZ, \calD, \calM, \calS, \bm T, \bm n)$ where
\begin{itemize}
  \item The multisets $\calZ, \calD, \calM$, and $\calS$ are over the
  base sets $\Sigma_{\out}$, $\Delta$, $\mon$, and $\sub$, respectively.
  These multisets will maintain the invariant that identity~\eqref{eqn:identity:no:emptyset} holds,
  as we shall show in the next section.
  \item $\bm n$ is a collection of non-negative real numbers (logs of statistics), one
  statistics $n_\delta$ for each $\delta \in \calD$.
  (Recall that, $n_\delta := \log N_\delta$, for convenience.)
  \item $\bm T$ is a collection of dictionaries, one
  dictionary $T_\delta$ for each $\delta \in \calD$; these shall be the guards
  for $N_\delta = 2^{n_\delta}$. In particular, $T_\delta\models N_\delta$ for each $\delta \in \calD$.
\end{itemize}

\begin{definition}[Terminal leaf]
    \label{defn:terminal:leaf}
A leaf node of the tree is ``terminal'' (it won't spawn off new sub-problem(s) anymore)
if its parameters
$(\calZ, \calD, \calM, \calS, \bm T, \bm n)$ are such that, there is an unconditional
statistics term $(\bm Z | \emptyset) \in \calD$ for which $\bm Z \in \calZ$.
\end{definition}

We next briefly explain in words the key steps of Algorithm~\ref{algo:panda}. A proof that
the algorithm is correct and an analysis of its runtime to show Theorem~\ref{th:algorithm}
are described in Section~\ref{subsec:analysis}.
The main loop looks for a non-terminal leaf node $\beta$ of the sub-problem tree. If there
is no such leaf node, then the code-block starting at line~\ref{line:gathering} gathers
the results from the parameters of the (terminal) leaf nodes to construct the final output
$\Sigma_\out$.

If there is a non-terminal leaf node $\beta$, then we pick an arbitrary
unconditional\footnote{We shall show that an unconditional $\delta$ exists in Section~\ref{subsec:analysis}.}
statistics term $\delta = (\bm W|\emptyset) \in \calD$
(line~\ref{line:arbitrary:delta}) and start considering cases in parallel with the cases in
the proof of Lemma~\ref{lemma:reset}, except that Case 3 will be handled differently. The major insight of our work is that we can design
an algorithm where the algorithmic steps mirror the Shannon inequality inference steps:
\begin{description}
  \item[Case 1: Join or Reset]
  There exists a statistics term $(\bm Y|\bm W) \in \calD$ (line~\ref{line:case1}).
  If the statistics are such that the join-size of $T_{\bm W} \Join T_{\bm Y|\bm W}$ is
  smaller than the budget $B_{\Delta, \bm N}$, checked at line~\ref{line:small:join}, then we compute
  the join result $T_{\bm Y\bm W}$ as shown, and let it guard the new statistics term $(\bm Y\bm W|\emptyset)$.
  This case corresponds precisely to applying Eq.~\eqref{eqn:D:step}, and the new multiset
  $\calD'$ defined in line~\ref{line:new:D'} reflects that.
  After joining $T_{\bm W}$ and $T_{\bm Y|\bm W}$ to form $T_{\bm Y\bm W}$, we remove the old
  dictionaries from $\bm T$ and add the new dictionary to $\bm T$. Similarly, the statistics
  in $\bm n$ are replaced in the same way.

  On the other hand, if the join $T_{\bm W} \Join T_{\bm Y|\bm W}$ is larger than the budget,
  then we perform a {\em reset} step. The Shannon inequality reasoning is as follows.
  Initially $\sum_{\bm Z \in \calZ}h(\bm Z) \leq \sum_{\delta \in \calD} h(\delta)$ holds,
  with witness terms $\calM$ and $\calS$. After applying the change in Eq.~\eqref{eqn:D:step},
  inequality $\sum_{\bm Z \in \calZ}h(\bm Z) \leq \sum_{\delta \in \overline \calD} h(\delta)$
  is still a Shannon inequality with the same witness. (Note that $\overline \calD$ is
  defined in line~\ref{line:overline:calD}). Now, we apply Lemma~\ref{lemma:reset} to {\em drop}
  $\delta_0 = (\bm Y\bm W | \emptyset)$ from $\overline\calD$, obtaining new parameters
  $(\calZ',\calD',\calM',\calS')$ to make progress on our algorithm.
  We remove from $\bm T$ and $\bm n$ the terms that were dropped from $\calD$.

  \item[Case 2: Projection]  There exists a monotonicity term $(\bm Y|\bm X) \in \calM$ with $\bm W = \bm X \bm Y$ (line~\ref{line:case2}).
  Here, the Shannon inequality is modified by applying the monotonicity-statistics
  cancellation in Eq.~\eqref{eqn:M:step}. This change is reflected in $\calM'$ and $\calD'$.
  The new statistics term $(\bm X | \emptyset) \in \calD'$ has a guard which is
  the projection $T_{\bm X}$.

  \item[Case 3: Partition]  There exists a submodularity measure $(\bm Y;\bm Z|\bm X) \in \calS$ with
  $\bm W = \bm X\bm Y$ (line~\ref{line:case3}). In this case, instead of applying identity~\eqref{eqn:S:step} from Lemma~\ref{lemma:reset}, we apply the following identity
  to maintain the Shannon inequality:
  \begin{align}
    h(\bm W) -h(\bm Y;\bm Z|\bm X)&=
    h(\bm W) -h(\bm X\bm Y) - h(\bm X\bm Z) + h(\bm X) + h(\bm X \bm Y \bm Z) \nonumber \\
    &= h(\bm X)  + h(\bm Y|\bm X \bm Z)
    \label{eqn:S:step:algo}
  \end{align}
  Identity~\eqref{eqn:S:step:algo} says that we have two new statistics terms, $h(\bm X|\emptyset)$ and
  $h(\bm Y | \bm X \bm Z)$, which need guards; this is where we will need to create a logarithmic
  number of sub-problems in order to create the guards of the right statistics.

  In particular, the algorithm performs a ``partitioning step'': it {\em uniformizes}
  $T_{\bm W} (= T_{\bm X\bm Y})$ by applying Lemma~\ref{lmm:partition}.
  Each of the sub-problems has the corresponding statistics as defined
  in the for-loop starting at line~\ref{line:sub:problem}.
  The table $T^i$ is on variables $\bm W=\bm X\bm Y$. Thus, in order to have it guard
  the new term $(\bm Y|\bm X\bm Z)$, we will need to apply a dictionary extension to it (see Section~\ref{subsec:tables:dictionaries}).
  Note that~\eqref{eq:partition} guarantees that
  $n^i_{\bm X} + n^i_{\bm Y|\bm X\bm Z} \leq n_{\bm X \bm Y} = n_{\bm W}$.
\end{description}

\subsection{Proof of correctness and runtime analysis}
\label{subsec:analysis}

In this section, we prove that the algorithm is correct and analyzes its runtime, to
complete the proof of Theorem~\ref{th:algorithm}.
Both of these tasks are accomplished by showing that \panda maintains the invariants
described below.

Recall that every sub-problem in the tree is parameterized by a tuple
$(\calZ, \calD, \calM, \calS, \bm T, \bm n)$.
For any sub-problem $\beta$ that is parameterized by
$(\calZ, \calD, \calM, \calS, \bm T, \bm n)$, define
\[ J_\beta := \dom^{\bm V} \Join \bigjoin_{\delta \in \calD}T_{\delta} \]
to be the join of all its dictionary guards.
At time $t$ in the algorithm, let $\calL_t$ denote the set of current leaf nodes in the
sub-problem tree.
While spawning new sub-problems, (we will show that)
\panda maintains the following invariants over all tuples $(\calZ, \calD, \calM, \calS, \bm T, \bm n)$: (Recall that $N_\delta \defeq 2^{n_\delta}$ for all $\delta \in \calD$.)

\begin{align}
  \mbox{Shannon inequality:} && \mbox{Identity~\eqref{eqn:identity:no:emptyset} } & \mbox{holds
w.r.t. the tuple $(\calZ, \calD, \calM, \calS)$} \label{eq:shannon:invariant}\\
  \mbox{Non-empty output:} && \calZ & \neq \emptyset \label{eq:invariant:output}\\
  \mbox{Lossless join:} &&  \bigjoin \Sigma_{\inn} & \subseteq
  \bigcup_{\beta\in \calL_t} J_\beta
  && \forall t \label{eq:invariant:join}\\
  \mbox{Small tables:} &&
    n_{\bm Y} &\leq b_{\Delta,\bm n}, && \forall (\bm Y| \emptyset) \in \calD \label{eq:invariant:small}\\
  \mbox{Guarded statistics:} && T_\delta &\models N_\delta, && \forall \delta \in \calD \label{eq:invariant:stats}\\
  \mbox{Upper bound:} && \frac{1}{|\calZ|}\sum_{\delta\in \calD} n_{\delta} &\leq b_{\Delta, \bm n}\label{eq:invariant:upper:bound}
\end{align}

We start by showing that, if the invariants hold, then the answer is correct with the
desired runtime.

\begin{proposition}
Let $N \defeq \max_{\delta \in \Delta} N_\delta$.
Suppose the invariants above are satisfied, then \panda returns a feasible solution
to the input disjunctive datalog rule in time $O(B_{\Delta, \bm N} \cdot (\log N)^{|\calS|})$,
where $\calS$ is the {\em input} multiset of submodularity measures.
\label{prop:correctness}
\end{proposition}
\begin{proof}
  For a given non-terminal leaf $\beta$, the number of unconditional statistics terms
  in $\calD$ is at least $|\calZ|$. This follows from Proposition~\ref{prop:whole:terms}
  and invariant~\eqref{eq:shannon:invariant}. Invariant~\eqref{eq:invariant:output} ensures
  that $|\calZ| \geq 1$, and thus the unconditional statistics term $\delta$ exists
  for line~\ref{line:arbitrary:delta} of the algorithm to proceed.
  Thanks to invariant~\eqref{eq:shannon:invariant},
  at least one of the three cases considered in the main loop must hold.
  Invariant~\eqref{eq:invariant:stats} guarantees that we can proceed to perform the join
  or the partition steps using corresponding tables / dictionaries when we need them
  in the algorithm.
  In summary, the body of the main loop can proceed as explained without getting stuck
  somewhere.

  For every node in the sub-problem tree that the algorithm creates, with parameter tuple
  $(\calZ, \calD, \calM, \calS, \bm T, \bm n)$, define the ``potential'' quantity
  $|\calD|+|\calM|+2|\calS|$ as in the proof of Lemma~\ref{lemma:reset}.
  Then, similar to what happens in the lemma's proof, the potential of the child is at least
  $1$ less than the potential of the parent.\footnote{Unlike Case 3 of Lemma~\ref{lemma:reset}
  where applying Eq.~\eqref{eqn:S:step} reduces $|\calS|$ by one and increases $|\calM|$ by one,
  in Case 3 of the algorithm, we apply Eq.~\eqref{eqn:S:step:algo} which reduces $|\calS|$ by one and increases $|\calD|$ by one.}\footnote{Note that the {\tt else} branch of Case 1 (line~\ref{line:else}) also reduces the potential by at least one, and applying the reset lemma in line~\ref{line:else:reset} never increases the potential.} Thus, the depth of the sub-problem tree
  is bounded by the original potential $|\calD|+|\calM|+2|\calS|$.
  This proves that the algorithm terminates.

  The time spent within each node $\beta$ of the tree is dominated by either
  the join step $T_{\bm Y\bm W}:=T_{\bm W}\Join T_{\bm Y | \bm W}$ in Case 1,
  the projection $T_{\bm X} := \pi_{\bm X}(T_{\bm W})$ in Case 2,
  or the partition step in Case 3 whose cost is $O(|T_{\bm W}|)$.
  In Case 2 and Case 3, the cost is bounded by $B_{\Delta,\bm N}$, thanks to
  invariant~\eqref{eq:invariant:small}.
  In Case 1, the join is only computed if $n_{\bm W}+n_{\bm Y | \bm W}\leq b_{\Delta,\bm n}$,
  thus it is also within the budget of $B_{\Delta,\bm N}$.
  Overall, the total time spent on all the nodes of the tree is bounded by $B_{\Delta,\bm N}$
  times the number of nodes. As we observed above, the depth of the tree is at most
  $|\calD|+|\calM|+2|\calS|$. Every node has a fanout of either $1$, or
  $k = O(\log B_{\Delta,\bm N}) = O(\log N)$.
  Every time the fanout is more than $1$, the number of submodularity measures in
  $\calS$ is reduced by $1$. Thus, the total runtime in data complexity is
  $O(B_{\Delta, \bm N} \cdot (\log N)^{|\calS|})$.

  Last but not least, we prove that the answer computed starting at line~\ref{line:gathering}
  is correct, which means according to Definition~\ref{def:feasible} that for every tuple $\bm t \in \bigjoin \Sigma_\inn$, there must
  exist $Q(\bm Z) \in \Sigma_\out$ for which $\pi_{\bm Z}(\bm t) \in Q(\bm Z)$.
  Note that Definition~\ref{def:feasible} does {\em not} require the converse to hold.\footnote{The converse will only become relevant in Section~\ref{sec:subw} when we discuss the evaluation of conjunctive queries.}
  In particular, the output may contain tuples that are not in $\bigjoin \Sigma_\inn$.
  In order to show that for every tuple $\bm t \in \bigjoin \Sigma_\inn$, there
  exists $Q(\bm Z) \in \Sigma_\out$ for which $\pi_{\bm Z}(\bm t) \in Q(\bm Z)$,
  we rely on invariant~\eqref{eq:invariant:join}.
  In particular, let $\calL$ denote the set of all final leaf nodes in the sub-problem tree.
  Then, every $\bm t \in \bigjoin\Sigma_\inn$ belongs to $J_\beta$ for some $\beta \in \calL$.
  Since $\beta$ is a terminal leaf node, there must exist $\delta = (\bm Z|\emptyset) \in \calD$
  such that $Q(\bm Z) \in \Sigma_\out$, and so $\pi_{\bm Z}(\bm t) \in Q(\bm Z)$
  thanks to line~\ref{line:got:it} of Algorithm~\ref{algo:panda}.
\end{proof}

\begin{proposition}
Algorithm~\ref{algo:panda} maintains
invariants~\eqref{eq:shannon:invariant} to \eqref{eq:invariant:upper:bound} throughout its
execution.
\label{prop:panda:invariants}
\end{proposition}
\begin{proof}
  We verify that every invariant from~\eqref{eq:shannon:invariant}
  to~\eqref{eq:invariant:upper:bound} holds one by one, by induction. For the input, only
  invariant~\eqref{eq:invariant:small} may not hold, because some input tables may be larger than
  the desired bound $B_{\Delta,\bm N}$. We deal with this situation by repeatedly applying
  the reset lemma  as was done in the {\tt else} branch of Case 1 (line~\ref{line:else}), dropping input tables
  that are too large. After this pre-processing step to make sure that all invariants are
  satisfied initially, we verify that they remain satisfied by induction.

  Invariant~\eqref{eq:shannon:invariant} is guaranteed by constructing the multisets
  $(\calZ', \calD', \calM', \calS')$ for the sub-problems while keeping~\eqref{eqn:identity:no:emptyset}
  intact. This is easy to verify in all three cases as we apply Eq.~\eqref{eqn:D:step},
  ~\eqref{eqn:M:step},~\eqref{eqn:S:step:algo} or Lemma~\ref{lemma:reset}.

  For invariant~\ref{eq:invariant:output}, the only place where $\calZ$ is changed is
  at line~\ref{line:calZ:2}. This happens when $n_{\bm W}+n_{\bm Y|\bm W}>b_{\Delta,\bm n}$.
  Since inequality~\eqref{eq:invariant:upper:bound} holds (at the previous iteration),
  we have
  \[
    b_{\Delta,\bm n}
    \geq \frac{1}{|\calZ|}\sum_{\delta\in\calD}n_\delta
    \geq \frac{1}{|\calZ|}(n_{\bm W}+n_{\bm Y|\bm W})
    > \frac{1}{|\calZ|} \cdot b_{\Delta,\bm n}
  \]
  It follows that $|\calZ| \geq 2$. Thus, when applying Lemma~\ref{lemma:reset}, we end up
  with $|\calZ'| \geq |\calZ|-1 \geq 1$, which means $\calZ'$ is not empty.

  Invariants~\eqref{eq:invariant:join},~\eqref{eq:invariant:small},~\eqref{eq:invariant:stats},
  and~\eqref{eq:invariant:upper:bound} can be verified one by one by simple case analysis.
  Below, we highlight the most prominent cases:
  \begin{itemize}
    \item Case 1 of the algorithm maintains invariant~\eqref{eq:invariant:small} specifically
    because we only do the join $T_{\bm Y\bm W} := T_{\bm W} \Join T_{\bm Y|\bm X}$
    when $n_{\bm Y\bm W} := n_{\bm W}+n_{\bm Y|\bm W} \leq b_{\Delta,\bm n}$.
    \item The {\tt else} branch of Case 1 (line~\ref{line:else}) maintains invariant~\eqref{eq:invariant:upper:bound} because of the following:
    \begin{align*}
        \sum_{\delta \in \calD'}n'_{\delta} = \sum_{\delta \in \calD'}n_{\delta} &\leq
        \sum_{\delta \in \calD}n_{\delta} - n_{\bm W} - n_{\bm Y|\bm W}&\text{(by Lemma~\ref{lemma:reset})}\\
        &<\sum_{\delta \in \calD}n_{\delta} - b_{\Delta,\bm n}&\text{($n_{\bm W} + n_{\bm Y\bm W} > b_{\Delta, \bm n}$)}\\
        &\leq |\calZ|\cdot b_{\Delta,\bm n} - b_{\Delta,\bm n}&\text{(by invariant~\eqref{eq:invariant:upper:bound} before)}\\
        &\leq |\calZ'|\cdot b_{\Delta,\bm n}&\text{(because $|\calZ'| \geq |\calZ| - 1$)}
    \end{align*}
    \item Case 3 of the algorithm maintains invariant~\eqref{eq:invariant:upper:bound}
    because Eq.~\eqref{eq:partition} implies that $n_{\bm X}^i + n_{\bm Y | \bm X \bm Z}^i \leq
    n_{\bm X\bm Y} = n_{\bm W}$. (Recall the definition of $n_{\bm X}^i$ and $n_{\bm Y | \bm X \bm Z}^i$ in line~\ref{line:sub:problem:stats}.)\qedhere
  \end{itemize}
\end{proof}

\section{Answering Conjunctive Queries in Submodular-Width Time}
\label{sec:subw}

The main aim of this section is to explain how \panda can be used to compute a conjunctive
query in time given by its submodular width (plus the output size). Recall the definition of
a conjunctive query in Equation~\eqref{eq:cq}.
In Marx's work~\cite{DBLP:journals/jacm/Marx13}, the submodular width was only defined for Boolean
conjunctive queries (i.e.,~$\bm F$ is empty) where all input relations are set to be of size
$N$. In Section~\ref{subsec:width:parameters}, we generalize this notion to the case when $\bm F$ is arbitrary and the
input relations satisfy given degree constraints, which are much more general than a single relation
size, and also subsume functional dependencies.
Marx~\cite{DBLP:journals/jacm/Marx13} gave an algorithm that can answer Boolean conjunctive queries in time
$O(f(|\bm V|)\cdot N^{c\cdot\subm(Q)})$ where $|\bm V|$ is the number of variables, $f$ is some computable function, $N$ is the input size, $c$ is a constant greater than $1$, and $\subm(Q)$
is the submodular width of $Q$.
Therefore, Marx's algorithm establishes fixed-parameter tractability of the class of Boolean conjunctive queries
where the submodular width $\subm(Q)$ is bounded.\footnote{Marx's work~\cite{DBLP:journals/jacm/Marx13} also proves the converse but only for {\em self-join-free} queries,
where a {\em self-join-free} query is a query whose body atoms have distinct relation symbols.
In particular, Marx's work proves that the class of self-join-free Boolean conjunctive queries
where the submodular width is unbounded is not fixed-parameter tractable, conditioned on the ETH conjecture.}
In contrast, in Section~\ref{subsec:panda:subw}, we present our algorithm, which in addition to handling arbitrary $\bm F$ and degree constraints, removes the constant $c$ from the runtime
at the expense of introducing a polylogarithmic factor in $N$.
In particular, our algorithm solves a Boolean conjunctive query in time
$O(f_1(|\bm V|)\cdot N^{\subm(Q)}\cdot (\log N)^{f_2(|\bm V|)})$,
where $f_1$ and $f_2$ are two computable functions.
Due to the extra term $(\log N)^{f_2(|\bm V|)}$, our algorithm cannot be used to establish
fixed-parameter tractability of the class of Boolean conjunctive queries with bounded submodular width.
However, it is more useful than Marx's algorithm when considering the {\em fine-grained complexity} of these queries.

\subsection{Width parameters for conjunctive queries under degree constraints}
\label{subsec:width:parameters}

Given a conjunctive query $Q$ in the form~\eqref{eq:cq}:
\[ Q(\bm F) \cd  \bigwedge_{R(\bm X)\in \Sigma}R(\bm X), \]
let $\calT(Q)$ denote the set of all free-connex tree decompositions of the query.\footnote{
For a fixed query $Q$, there are at most $2^{2^{|\bm V|}}$ tree decompositions,
since any two trees that have the same set of bags can be considered
equal for our purposes.}
(See Section~\ref{subsec:td} for the definition of a free-connex tree decomposition.)
Let $(\Delta,\bm N)$ be a set of degree constraints, as defined in
Sec.~\ref{subsec:degrees}. As usual, we denote by $(\Delta, \bm n)$
the associated log-degree constraints.  We say that a polymatroid
$\bm h$ {\em satisfies} the constraints, and write
$\bm h \models (\Delta, \bm n)$, if $h(\delta) \leq n_\delta$ for all
$\delta \in \Delta$.

\begin{definition} \label{def:subw} The {\em degree-aware fractional
    hypertree width} and the {\em degree-aware submodular width} of
  $Q$ under the degree constraints $(\Delta, \bm n)$ are:
  \begin{align}
    \fhtw(Q,\Delta,\bm n) \defeq & \min_{(T,\chi)\in \calT(Q)} \max_{\bm h\models(\Delta,\bm n)} \max_{t \in \nodes(T)}h(\chi(t)) \label{eq:fhtw} \\
    \subm(Q,\Delta,\bm n) \defeq & \max_{\bm h\models(\Delta,\bm n)} \min_{(T,\chi)\in \calT(Q)} \max_{t \in \nodes(T)}h(\chi(t)) \label{eq:subm}
  \end{align}
(Note that $\bm h \models (\Delta,\bm n)$ requires $\bm h$ to both be a polymatroid
and satisfy the degree constraints.)
\end{definition}
\begin{remark}
Eq.~\eqref{eq:fhtw} and~\eqref{eq:subm} collapse back to the standard definitions of the fractional
hypertree width~\cite{DBLP:journals/talg/GroheM14} and submodular width~\cite{DBLP:journals/jacm/Marx13}, respectively, when the degree constraints $(\Delta, \bm N)$
only contain cardinality constraints of the form $|T_{\bm Y}| \leq N$ for a single number $N$ that represents the input size;
see~\cite{DBLP:conf/pods/Khamis0S17} for a proof.
Note, however, that in the standard definitions of $\fhtw$ and $\subm$,
the base of the log function was $N$, the input size, and thus runtimes were stated in the form
$O(N^{\fhtw})$ and $O(N^{\subm})$.  In our generalization, the base of the log function is
$2$, and thus runtimes are stated in the form $O(2^{\fhtw})$ and $O(2^{\subm})$.

It is straightforward to use \panda to answer a conjunctive query in time $\tilde
O(\norm{\Sigma}+ 2^{\fhtw})$, but we will only describe the algorithm for the
submodular width. The key advantage of $\fhtw$ over $\subm$ is that we can answer sum-product
queries over any semiring in time $\tilde O(\norm{\Sigma}+ 2^{\fhtw})$ using variable
elimination~\cite{DBLP:conf/pods/KhamisNR16}, since the query plan involves only {\em one}
(optimal) tree decomposition.

We will also show how to use $\panda$ to support {\em constant-delay enumeration}\footnote{By that, we mean reporting the output tuples one by one
where the time needed to report the next tuple (or report that none exists) remains a constant
throughout the entire process. The constant here is in {\em data complexity}.} of the output of a conjunctive query
after a preprocessing time of $\tilde O(\norm{\Sigma}+ 2^{\subm})$.\footnote{In particular, within the mentioned preprocessing time,
    we can produce a constant number of tree decompositions
    that cover the full join of the input relations.
    However, the same output tuple might be duplicated across multiple tree decompositions.
    Nevertheless, we can use the {\em Cheater's Lemma}~\cite{10.1145/3450263} to deduplicate
    the output while maintaining constant delay. See the proof of Theorem~\ref{thm:algo:subw} for more details.
}
Another line of work~\cite{berkholz_et_al:LIPIcs.MFCS.2019.58} shows how to support constant-delay enumeration after a preprocessing time of $\tilde O(\norm{\Sigma}+ 2^{c\cdot\subm})$
for some constant $c > 1$.
\end{remark}

The two definitions~\eqref{eq:fhtw} and~\eqref{eq:subm} differ only in the first two operations:
$\min \max$ v.s. $\max \min$.
It is easy to see that $\subm \leq \fhtw$, as it follows immediately from
the max-min inequality, which states that
$\max_x\min_y f(x, y) \leq \min_y\max_x f(x, y)$
for any function $f$.
As mentioned above, the degree-aware submodular width generalizes the submodular
width considering richer sets of statistics on the input data.  The original
definition assumed only cardinality constraints: there is one
cardinality constraint for each input relation, and they are all
equal.  In that case, both $\subm$ and $\fhtw$ are finite.  In our
generalization, $\subm$ can be $\infty$, for example when
$\Delta=\emptyset$.  When no confusion arises, we will simply call
$\fhtw$ and $\subm$ the fractional hypertree width and submodular width,
dropping the term {\em degree-aware}.

\begin{example}[$\fhtw$ and $\subm$ of the $k$-cycle with $k \geq 4$]
    \label{ex:k-cycle}
Consider the Boolean $k$-cycle query with $k \geq 4$:
\begin{align*}
    Q() &\cd R_{1,2}(X_1,X_2) \wedge \cdots \wedge R_{k-1,k}(X_{k-1},X_k) \wedge R_{k,1}(X_k,X_1).
\end{align*}
Suppose we have input cardinality statistics $N:=|R_{1,2}|=\cdots=|R_{k-1,k}|=|R_{k,1}|$. For instance, in the query
where all input relations are the edge set of a graph, $N$ is the number of edges.
We will show that $\subm \leq (2 - 1/\lceil k/2 \rceil )\log N$ and $\fhtw \geq 2\log N$ for this query.

To show the bound on the $\fhtw$, note that, in any tree decomposition $(T,\chi)$, there must be
at least one bag $\chi(t)$ that contains some three consecutive variables $\{X_{i-1},X_i,X_{i+1}\}$
on the cycle. Fix $i$ accordingly and consider the polymatroid $\bm h$ defined by:
\begin{align*}
h(\bm X) = |\bm X \cap \{X_{i-1},X_{i+1}\}|\cdot \log N && \text{for all $\bm X \subseteq \{X_1,\dots,X_k\}$}
\end{align*}
It is straightforward to verify that this is a polymatroid with $h(\chi(t))=2\log N$ and
$\bm h \models (\Delta, \bm n)$.

To prove the bound on $\subm$, consider any polymatroid $\bm h \models (\Delta,\bm n)$.
Let $\theta$ be a parameter to be determined. Consider two cases.
\begin{itemize}
    \item There exists $X_i$ for which $h(X_i)\leq \theta$. Without loss of generality, assume
    $h(X_1)\leq \theta$. Consider the tree decomposition:
    
\medskip 

 \includegraphics[width=0.75\textwidth]{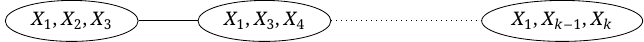}

      For every bag $B=\{ X_1, X_i, X_{i+1}\}$, we have
      $h(B) \leq h(X_1)+h(X_iX_{i+1}) \leq \theta+ \log N$.
    \item $h(X_i)>\theta$ for all $i \in [k]$. Consider the tree
      decomposition
      
      \medskip 

\includegraphics[width=0.75\textwidth]{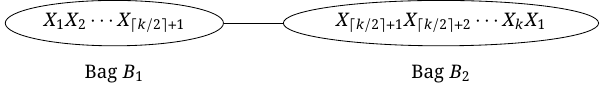}

   From submodularity,
   \begin{eqnarray*}
      h(B_1) \leq & h(X_1X_2) + \displaystyle{\sum_{i=3}^{\lceil k/2\rceil +1}h(X_i|X_{i-1})}
      & \leq \lceil k/2\rceil \log N - (\lceil k/2 \rceil -1) \theta \\
      h(B_2) \leq & h(X_{k}X_{1}) + \displaystyle{\sum_{i=\lceil k/2\rceil+1}^{k-1}h(X_i|X_{i+1})}
      & \leq \lfloor k/2\rfloor \log N - (\lfloor k/2 \rfloor -1) \theta.
   \end{eqnarray*}
   Setting $\theta = (1-1/\lceil k/2 \rceil)\log N$ to balance the two cases, we conclude that
   $\subm \leq (2-1/\lceil k/2 \rceil ) \log N$. \qedhere
\end{itemize}
\end{example}

\subsection{Achieving submodular width runtime with \panda}
\label{subsec:panda:subw}

Before explaining how \panda can be used to achieve the submodular width runtime,
we need a technical lemma.
\begin{lemma}
  \label{lmm:technical}
  Let $\calZ$ denote a finite collection\footnote{Note that $\calZ$ is {\em not} a multiset here.} of subsets of $\bm V$.
  Let $(\Delta,\bm n)$ denote given input degree constraints.
  If the following quantity is finite:
  \begin{align}
    \opt := \max_{\bm h \models (\Delta,\bm n)} \min_{\bm Z \in \calZ} h(\bm Z),
    \label{eq:max:min:problem}
  \end{align}
  then we can compute coefficients
  $\bm\lambda = (\lambda_{\bm Z})_{{\bm Z} \in \calZ}$ and
  $\bm w = (w_{\delta})_{\delta \in \Delta}$ such that the following are
  satisfied:
  \begin{itemize}
    \item[(a)] $\norm{\bm \lambda}_1=1$, $\bm \lambda \geq \bm 0$, and $\bm w \geq \bm 0$,
    \item[(b)]
  Inequality $\sum_{{\bm Z} \in \calZ} \lambda_{\bm Z} \cdot h({\bm Z}) \leq \sum_{\delta \in \Delta} w_\delta \cdot h(\delta)$
  is a Shannon inequality,
    \item[(c)] $\opt = \sum_{\delta\in\Delta}w_\delta n_\delta$.
  \end{itemize}
\end{lemma}
\begin{proof}
  Let $\Gamma$ denote the (polyhedral) set of all polymatroids over $\bm V$.
  We write $\opt$ in a slightly different form, where we introduce a new unconstrained variable
  $t$ to replace the inner $\min$:
  \begin{align} \opt = \max_{t, \bm h \in \Gamma}
      \{
        t \mid
        \forall {\bm Z} \in \calZ : t \leq h({\bm Z}),
        \text{ and }
        \forall \delta \in \Delta : h(\delta) \leq n_\delta
      \}
      \label{eq:max:min:problem:lp}
  \end{align}
  Introduce a Lagrangian multiplier $\lambda_{\bm Z}$ for every constraint $t \leq h({\bm Z})$,
  and $w_\delta$ for every constraint $h(\delta) \leq n_\delta$. The Lagrange dual
  function is
  \begin{align*}
      \calL(\bm \lambda, \bm w)
      &= \max_{t, \bm h \in \Gamma}
        \left( t + \sum_{{\bm Z} \in \calZ}\lambda_{\bm Z} (h({\bm Z})-t)
        + \sum_{\delta \in \Delta} w_\delta (n_\delta-h(\delta))
        \right) \\
      &= \sum_{\delta\in\Delta}w_\delta n_\delta
      + \max_{t} (1-\norm{\bm\lambda}_1)t
      + \max_{\bm h \in \Gamma} \left(
          \sum_{{\bm Z} \in \calZ} \lambda_{\bm Z}h({\bm Z}) -
          \sum_{\delta \in \Delta} w_\delta h(\delta)
        \right)
  \end{align*}
Let $\bm \lambda^*$ and $\bm w^*$ denote an optimal solution to the Lagrangian dual
problem
$\min \{ \calL(\bm \lambda, \bm w) \mid \bm\lambda \geq \bm 0, \bm w \geq \bm 0 \}$, then
by strong duality\footnote{Which holds because the problem is linear.}
\[
  \opt = \calL(\bm\lambda^*, \bm w^*)
      = \sum_{\delta\in\Delta}w^*_\delta n_\delta
      + \max_{t} (1-\norm{\bm\lambda^*}_1)t
      + \max_{\bm h \in \Gamma} \left(
          \sum_{{\bm Z} \in \calZ} \lambda^*_{\bm Z}h({\bm Z}) - \sum_{\delta \in \Delta} w^*_\delta h(\delta)
        \right)
\]
From the assumption that $\opt$ is finite, it follows that $\norm{\bm\lambda^*}_1=1$ because $t$
is unconstrained.
Furthermore, if there is any polymatroid $\bm h$ for which
$\sum_{{\bm Z} \in \calZ} \lambda^*_{\bm Z}h({\bm Z}) - \sum_{\delta \in \Delta} w^*_\delta h(\delta)
>0$ then
$\calL(\bm\lambda^*, \bm w^*)$ is unbounded, because any positive multiple of a polymatroid
is a polymatroid.
Thus, $(b)$ is satisfied. Furthermore, as the expression inside $\max_{\bm h \in \Gamma}$
is non-positive, the maximum it can achieve is $0$ with $\bm h = \bm 0$.
Consequently, $\bm\lambda^*$ and $\bm w^*$ satisfy the three conditions (a), (b), and (c)
above, and we can compute them with standard linear programming algorithms.
\end{proof}

Equipped with this tool, we are now ready to show how \panda can be used to answer a
conjunctive query in submodular width time:

\begin{theorem}  Given a set of degree constraints
  $(\Delta, \bm N)$, a conjunctive query $Q$ of the form~\eqref{eq:cq} can be computed in
  time
  \[ \tilde O(\norm{\Sigma} + 2^{\subm(Q, \Delta, \bm n)} + |\text{output}|) \]
  on any database instance $\Sigma$ that satisfies the degree constraints.
  In particular, within a preprocessing time of $\tilde O(\norm{\Sigma} + 2^{\subm(Q, \Delta, \bm n)})$,
  we can support constant-delay enumeration of the output.
  \label{thm:algo:subw}
\end{theorem}
\begin{proof}
  Let $\calT(Q)=\set{(T_1,\chi_1), \ldots, (T_m, \chi_m)}$ be all free-connex tree
  decompositions of $Q$. For every tree decomposition $(T_i, \chi_i) \in \calT(Q)$ and every
  node $j \in \nodes(T_i)$, create a fresh atom $A_{ij}(\bm Z_{ij})$ over variables $\bm
  Z_{ij} := \chi_i(j)$. In other words, every bag of every tree decomposition is associated
  with an atom. Let $\Sigma^i := \{ A_{ij}(\bm Z_{ij}) \mid j \in \nodes(T_i)
  \}$ denote a schema corresponding to the bags of the $i$th tree decomposition. The
  algorithm will compute relation instances $A_{ij}(\bm Z_{ij})$, for all $i \in [m]$ and $j
  \in \nodes(T_i)$, such that the tree decompositions together cover the full join of the
  input relations:
  \begin{align}
    \bigjoin \Sigma \subseteq \bigcup_{i \in [m]} \bigjoin \Sigma^i
    \label{eqn:main:invariant}
  \end{align}
  From these instances, there are $m$ separate free-connex acyclic conjunctive
  queries of the form
  \[Q^i(\bm F) \cd \bigwedge_{A_{ij}(\bm Z_{ij}) \in \Sigma^i}A_{ij}(\bm Z_{ij}),\] which
  can be computed in $\tilde O(\norm{\Sigma^i} + |Q^i(\bm F)|)$ time, using
  Lemma~\ref{lmm:free:connex:acyclic}.
  Before computing these queries, we semijoin reduce each $A_{ij}(\bm Z_{ij})$ in $\Sigma^i$
  with all the input relations in $\Sigma$.
  Recall that by definition of a tree decomposition, every input relation $R(\bm X)$ in $\Sigma$
  must have its variables $\bm X$ appear in some bag of the tree decomposition $(T_i, \chi_i)$, hence
  in some $\bm Z_{ij}$ in $\Sigma^i$.
  Therefore, this semijoin reduction ensures that the output of each query $Q^i(\bm F)$ is a subset of the full join of the input relations $\bigjoin \Sigma$. This turns~\eqref{eqn:main:invariant} into an equality.

To support constant-delay enumeration of the output after a preprocessing time of $\tilde O(\norm{\Sigma} + 2^{\subm(Q, \Delta, \bm n)})$,
we will use constant-delay enumeration of the output of each $Q^i(\bm F)$
after a preprocessing time of $\tilde O(\norm{\Sigma^i})$, as explained in Lemma~\ref{lmm:free:connex:acyclic}.
Note that the same output tuple might be duplicated across multiple queries $Q^i(\bm F)$.
Nevertheless, we can use the {\em Cheater's Lemma}~\cite{10.1145/3450263} to deduplicate the output while maintaining constant delay.\footnote{Recall that the number of tree decompositions is constant in data complexity.}

  It remains to show that we can compute all the instances $\Sigma^i$
  satisfying~\eqref{eqn:main:invariant} in time \linebreak $\tilde O(2^{\subm(Q, \Delta, \bm n)})$.
  Obviously, to compute them in time $\tilde O(2^{\subm(Q, \Delta, \bm n)})$, it is
  necessary that $\norm{\Sigma^i} = \tilde O(2^{\subm(Q, \Delta, \bm n)})$. To this
  end, for every combination of nodes $\bm j = (j_1, j_2, \dots, j_m) \in \nodes(T_1) \times
  \cdots \times \nodes(T_m)$, we will compute a feasible output $B^{\bm j}_1, \dots B^{\bm
  j}_m$ to the following DDR (whose input schema is $\Sigma$):
  \begin{align}
     \bigvee_{i \in [m]} B^{\bm j}_i(\bm Z_{ij_i}) & \cd \bigwedge_{R(\bm X) \in \Sigma}R(\bm X).
     && \text{(the $\bm j$th DDR)}
     \label{eqn:ddr:j}
  \end{align}
  In words, for this DDR, there is a representative bag $B_i^{\bm j}$ from each tree decomposition
  $(T_i,\chi_i)$.
  After feasible solutions to all these DDRs are computed, then we set
  $A_{ij} := \bigcup_{\bm j : j_i = j} B_i^{\bm j}.$

  We first prove that the instances $A_{ij}$ defined as such
  satisfy property~\eqref{eqn:main:invariant}. Suppose there is a tuple $\bm t \in
  \bigjoin \Sigma$ that is not in the RHS of~\eqref{eqn:main:invariant}. Then, for
  every $i \in [m]$, there exists a node $j_i \in \nodes(T_i)$ such that $\pi_{\bm
  Z_{ij_i}}(\bm t) \not\in A_{ij_i}$. Collect these $j_i$ into a tuple $\bm j$, then this
  implies that we did not compute a feasible output to the $\bm j$th
  DDR~\eqref{eqn:ddr:j}, a contradiction.

  Last but not least, we show that the DDRs~\eqref{eqn:ddr:j} can be computed in time
  $\tilde O(2^{\subm(Q, \Delta, \bm n)})$.  Fix a tuple of nodes $\bm j = (j_1, \dots, j_m)$.
  Let $\calZ$ denote the set of all bags $\bm Z_{ij_i}$ for $i \in [m]$.
  Define
  \[ \opt := \max_{\bm h \models (\Delta, \bm n)} \min_{Z \in \calZ} h(Z).  \]
  Then,
  \begin{align*}
    \opt = \max_{\bm h \models (\Delta, \bm n)} \min_{i \in [m]} h(Z_{ij_i})
         \leq \max_{\bm h \models (\Delta, \bm n)} \min_{i \in [m]} \max_{t \in \nodes(T_i)} h(\chi(t))
         = \subm(Q, \Delta, \bm n).
  \end{align*}

  WLOG, we assume that $\subm$ is finite, which means $\opt$ is finite.
  By
  Lemma~\ref{lmm:technical}, we can compute coefficients $\bm \lambda$ and $\bm w$ such that
  the three conditions in Lemma~\ref{lmm:technical} are satisfied. Hence, from
  Theorem~\ref{th:algorithm}, we can compute a feasible output to the DDR~\eqref{eqn:ddr:j}
  in time $\tilde O(2^\opt) = \tilde O(2^{\subm(Q, \Delta, \bm n)})$.
\end{proof}

\subsection{Example: Solving a conjunctive query in submodular width time}
\begin{figure}[t]
        \centering
       \includegraphics[width=0.8\textwidth]{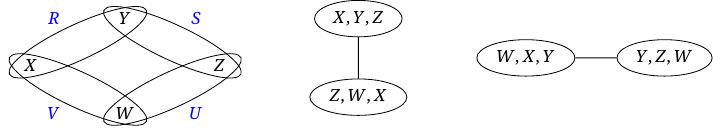}
    \caption{Query~\eqref{eq:4cycle} with the two free-connex tree decompositions.}
    \label{fig:4cycle}
    \end{figure}

    Consider the following query $Q$ whose body is a 4-cycle:
    \begin{align}
        Q(X, Y) \cd & R(X,Y) \wedge S(Y,Z) \wedge U(Z,W) \wedge V(W, X)\label{eq:4cycle}
    \end{align}
    Suppose we only have cardinality statistics where all input relation sizes are upper bounded by $N$ for some number $N$, i.e.,
    \begin{align*}
        \Delta = & \set{(XY|\emptyset), (YZ|\emptyset), (ZW|\emptyset), (WX|\emptyset)},
        & N_{XY}=&N_{YZ}=N_{ZW}=N_{WX}=N.
    \end{align*}
    Let $n := \log N$. This query has two free-connex
    tree decompositions, depicted in Figure~\ref{fig:4cycle} (ignoring the trivial tree decomposition with a single bag):
    \begin{itemize}
        \item One with two bags $A_{11}(X, Y, Z)$ and $A_{12}(Z, W, X)$.
        \item One with two bags $A_{21}(Y, Z, W)$ and $A_{22}(W, X, Y)$.
    \end{itemize}

    It is not hard to see that the degree-aware fractional hypertree width of $Q$, given by~\eqref{eq:fhtw}, is exactly $2 n$, and we leave this as an exercise. Next, we show that the degree-aware submodular width is $3n/2$. In particular, Eq.~\eqref{eq:subm} for this query becomes:
    \begin{align*}
        \subm(Q, \Delta, \bm n) = & \max_{\bm h\models (\Delta, \bm n)}
        \min(\max(h(XYZ),h(ZWX)), \max(h(YZW), h(WXY)))
    \end{align*}
    By distributing the $\min$ over the inner $\max$, and then swapping the two $\max$ operators, we get:
    \begin{align}
        \subm(Q, \Delta, \bm n) = \max(&\nonumber\\
            & \max_{\bm h \models (\Delta, \bm n)} \min(h(XYZ), h(YZW)), \label{eq:bag-selector:1}\\
            & \max_{\bm h \models (\Delta, \bm n)} \min(h(XYZ), h(WXY)), \label{eq:bag-selector:2}\\
            & \max_{\bm h \models (\Delta, \bm n)} \min(h(ZWX), h(YZW)), \label{eq:bag-selector:3}\\
            & \max_{\bm h \models (\Delta, \bm n)} \min(h(ZWX), h(WXY))) \label{eq:bag-selector:4}
    \end{align}
    Note that each of the expressions $\eqref{eq:bag-selector:1}$\ldots$\eqref{eq:bag-selector:4}$
    has the same format as the optimization problem~\eqref{eq:max:min:problem} in Lemma~\ref{lmm:technical} and is equivalent to the linear program~\eqref{eq:max:min:problem:lp}.
    Let's take the first expression~\eqref{eq:bag-selector:1}. (The other three are similar.)
    For this expression, a linear program solver gives $\opt = 3n/2$. Lemma~\eqref{lmm:technical} guarantees for us the existence of the following Shannon inequality, which is the same as~\eqref{eq:panda:example:shannon} from Section~\ref{subsec:example2}:
    \begin{align*}
        \frac 1 2 h(XYZ) + \frac 1 2 h(YZW) \leq & \frac 1 2 h(XY|\emptyset)+\frac 1 2 h(YZ|\emptyset)+\frac 1 2 h(ZW|\emptyset)
      \end{align*}
      In particular, by item (c) of the lemma, we have:
      \[
        \opt = \frac{1}{2} n_{XY} + \frac{1}{2} n_{YZ} + \frac{1}{2} n_{ZW} = \frac{3}{2} n.
      \]
    The other three expressions~\eqref{eq:bag-selector:2}--\eqref{eq:bag-selector:4} also have
    $\opt = 3 n/2$, leading to $\subm(Q, \Delta, \bm n) = 3 n/ 2$.

    Next, we describe the algorithm to compute the query~\eqref{eq:4cycle} in time
    $\tilde O(N^{3/2} + |\text{output}|)$, as claimed in Theorem~\ref{thm:algo:subw}.
    The algorithm starts by constructing the following four DDRs, which mirror the four
    expressions~\eqref{eq:bag-selector:1}--\eqref{eq:bag-selector:4}:
    \begin{align}
        B_1^{11}(X,Y,Z) \vee B_2^{11}(Y,Z,W) & \cd
            R(X,Y) \wedge S(Y,Z) \wedge U(Z,W) \wedge V(W,X) \label{eq:ddr1}\\
        B_1^{12}(X,Y,Z) \vee B_2^{12}(W,X,Y) & \cd
            R(X,Y) \wedge S(Y,Z) \wedge U(Z,W) \wedge V(W,X) \label{eq:ddr2}\\
        B_1^{21}(Z,W,X) \vee B_2^{21}(Y,Z,W) & \cd
            R(X,Y) \wedge S(Y,Z) \wedge U(Z,W) \wedge V(W,X) \label{eq:ddr3}\\
        B_1^{22}(Z,W,X) \vee B_2^{22}(W,X,Y) & \cd
            R(X,Y) \wedge S(Y,Z) \wedge U(Z,W) \wedge V(W,X) \label{eq:ddr4}
    \end{align}
    Let's take the first DDR~\eqref{eq:ddr1} as an example. We can compute a feasible output
    to this DDR by computing a feasible output to the following DDR instead:
    \begin{align*}
        B_1^{11}(X,Y,Z) \vee B_2^{11}(Y,Z,W) & \cd
            R(X,Y) \wedge S(Y,Z) \wedge U(Z,W)
    \end{align*}
    The above DDR is identical to~\eqref{eq:ab:query}, and we saw in Section~\ref{subsec:example2} that we can compute a feasible output to it in time $\tilde O(N^{3/2})$.
    The other 3 DDRs~\eqref{eq:ddr2}--\eqref{eq:ddr4} can be computed in the same way.
    Afterwards, we compute:
    \begin{align*}
        A_{11} & := B_1^{11} \cup B_1^{12} \\
        A_{12} & := B_1^{21} \cup B_1^{22} \\
        A_{21} & := B_2^{11} \cup B_2^{21} \\
        A_{22} & := B_2^{12} \cup B_2^{22}
    \end{align*}
    Using Lemma~\eqref{lmm:free:connex:acyclic}, we compute the following two free-connex acyclic conjunctive queries
    (after {\em semijoin-reducing} each of the $A_{ij}$ relations with the input relations $R, S, U$ and $V$):
    \begin{align*}
        Q^1(X, Y) & \cd A_{11}(X, Y, Z) \wedge A_{12}(Z, W, X) \\
        Q^2(X, Y) & \cd A_{21}(Y, Z, W) \wedge A_{22}(W, X, Y)
    \end{align*}
    Finally, we take the union of $Q^1$ and $Q^2$ above as the output $Q$ to the query in~\eqref{eq:4cycle}. The overall runtime is $\tilde O(N^{3/2} + |\text{output}|)$.

    In order to prove the correctness of this algorithm, we show that the full join $R(X, Y)\Join S(Y, Z)\Join U(Z, W)\Join V(W, X)$ is identical to the set of tuples $(x, y, z, w)$ that satisfy:
    \begin{align}
        (A_{11}(x, y, z) \wedge A_{12}(z, w, x)) \vee
        (A_{21}(y, z, w) \wedge A_{22}(w, x, y))
        \label{eq:correctness:condition}
    \end{align}
    The containment $\supseteq$ is immediate from the definition of $A_{ij}$ (and thanks to the semijoin reduction of $A_{ij}$ with the input relations $R, S, U$ and $V$). For the other containment $\subseteq$, note that condition~\eqref{eq:correctness:condition} is equivalent to the following:
    \begin{align}
        \label{eq:correctness:condition2}
        &(A_{11}(x, y, z) \vee A_{21}(y, z, w)) \wedge\\
        &(A_{11}(x, y, z) \vee A_{22}(w, x, y)) \wedge\nonumber\\
        &(A_{12}(z, w, x) \vee A_{21}(y, z, w)) \wedge\nonumber\\
        &(A_{12}(z, w, x) \vee A_{22}(w, x, y)) \nonumber
    \end{align}
    By~\eqref{eq:ddr1}\ldots~\eqref{eq:ddr4}, every tuple $(x, y, z, w)$ that satisfies the conjunction $R(x, y)\wedge S(y, z)\wedge U(z, w)\wedge V(w, x)$ must also satisfy~\eqref{eq:correctness:condition2}. This completes the proof of correctness.

\section{Conclusion}
\label{sec:conclusion}

We presented $\panda$, an algorithm that computes a conjunctive query in time given by its
submodular width.  For this purpose, we have used a generalization of the notion of
submodular width in~\cite{DBLP:journals/jacm/Marx13}, by incorporating a rich class of
statistics on the input relations, including cardinality constraints and degree constraints;
the latter can also express functional dependencies. The \panda algorithm described here is
a significant simplification of its preliminary version in~\cite{DBLP:conf/pods/Khamis0S17}.
\panda can also be used as a Worst-Case-Optimal-Join algorithm to compute the output of a
full conjunctive query in time bounded by the information-theoretic upper bound of the
output size. A recent extension showed that it can also be extended to account for
$\ell_p$-norms of degree sequences in the input~\cite{10.1145/3651597}.

We leave some open problems.  The first is an analysis of the
complexity of the witness of a Shannon inequality.  The number of
submodularities in the Shannon inequality appears as the exponent
of a logarithmic factor in the runtime of \panda, and it would be very useful to
study this number as a function of the query.
Another question concerns the
number of tree decompositions needed to compute the submodular width:
our current bound is double exponential, and the question is whether
this can be reduced.  Finally, one open problem is whether $\panda$
can be generalized to achieve information-theoretic bounds
corresponding to {\em non-Shannon
  inequalities}~\cite{DBLP:journals/tit/ZhangY97,zhang1998characterization}.

\printbibliography

\end{document}